\documentclass{aa}  
\usepackage{mathabx}
\usepackage{graphicx}
\usepackage{txfonts}
\usepackage{natbib}
\usepackage{booktabs}
\bibpunct{(}{)}{;}{a}{}{,} 
\usepackage{subfig}
\usepackage{hyperref}
\usepackage{xcolor}
\hypersetup{unicode}
\hypersetup{breaklinks=true}
\hypersetup{colorlinks,					
	linkcolor={red!50!black},
	citecolor={blue!50!black},
	urlcolor={blue!80!black}}

\newcommand{\orcid}[1]{\protect\href{https://orcid.org/#1}{\protect\includegraphics[width=8pt]{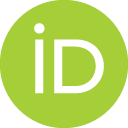}}}
\usepackage[normalem]{ulem}
\usepackage[markup=bfit]{changes}
\definechangesauthor[name={Hritam Chakraborty}, color=violet]{HC}
\definechangesauthor[name={Hritam Chakraborty}, color=violet]{hc}
\definechangesauthor[name={Arianna Nigioni}, color=blue]{an}
\usepackage[separate-uncertainty = true,multi-part-units=single,group-separator={}, group-digits = integer, group-four-digits = true,per-mode=reciprocal]{siunitx}
\DeclareSIUnit\parsec{pc}
\usepackage{chemformula}
\usepackage{threeparttable}

\begin{document}

   \title{Transit timing and starspot-induced transit depth variations in the $\sim$17 Myr old HIP 67522 system\thanks{This study uses CHEOPS data observed as part of the Guest Observers (GO) programmes CH\_PR240017 (PI Chakraborty) and CH\_PR240004 (PI Ilin).}}

   \author{H.~Chakraborty\inst{\ref{geneva}}\thanks{\href{mailto:Hritam.Chakraborty@unige.ch}{Email: Hritam.Chakraborty@unige.ch}}\orcid{0000-0002-5177-1898}
          \and       
          M.~Lendl\inst{\ref{geneva}}\orcid{0000-0001-9699-1459} 
         \and
         A.~Nigioni\inst{\ref{geneva}}\orcid{0009-0004-5882-6574}
         \and
         J.~A.~Egger\inst{\ref{bern},\ref{estec}}\thanks{ESA Research Fellow}\orcid{0000-0003-1628-4231}
         \and
         J.M.~Almenara\inst{\ref{geneva}}\orcid{0000-0003-3208-9815}
         \and
         M.~P.~Battley\inst{\ref{queen_mary}, \ref{geneva}}\orcid{0000-0002-1357-9774}
         \and
         A.~G.~M.~Pietrow\inst{\ref{potsdam}}\orcid{0000-0002-0484-7634}
         \and
         J.~Venturini\inst{\ref{geneva}}\orcid{0000-0001-9527-2903}
         \and
         A.~Heitzmann\orcid{0000-0002-8091-7526}\inst{\ref{geneva}}
         \and
         B.~Akinsanmi\inst{\ref{geneva}}\orcid{/0000-0001-6519-1598}
         \and
         I.~Apergis\inst{\ref{warwick1},\ref{warwick2}}\orcid{0009-0004-7473-4573}
         \and
         D.~Bayliss\inst{\ref{warwick1}, \ref{warwick2}}\orcid{0000-0001-6023-1335}
         \and
         J.~P.~Faria\inst{\ref{geneva}}\orcid{0000-0002-6728-244X}
         \and
         A.~D.~Feinstein\inst{\ref{michigan}}\thanks{NHFP Sagan Fellow}\orcid{0000-0002-9464-8101}
         \and
         J.~F.~Fern\'andez\inst{\ref{warwick1},\ref{warwick2}}\orcid{0000-0002-1416-2188}
         \and
         E.~Fontanet\inst{\ref{geneva}}\orcid{0000-0002-0215-4551}
         \and
         E.~Gillen\inst{\ref{queen_mary}}\orcid{0000-0003-2851-3070}
         \and
         E.~Ilin\orcid{0000-0002-6299-7542}\inst{\ref{astron}}
         \and
         T.~Jestin\orcid{0009-0000-5585-7915}\inst{\ref{geneva},\ref{epfl}}
         \and
         J.~S.~Jenkins\inst{\ref{chile1},\ref{chile2}}\orcid{0000-0003-2733-8725}
         \and
         E.~J.~Lößnitz\inst{\ref{potsdam}, \ref{potsdam2}}\orcid{0009-0006-3049-4875}
         \and
         J.~McCormac\inst{\ref{warwick1},\ref{warwick2}}\orcid{0000-0003-1631-4170}
         \and
         C.~Modrasini\inst{\ref{bern}}\orcid{0000-0002-1013-2811}
         \and
         F.~Neiradiaz\inst{\ref{epfl}}
         \and
         H.~P.~Osborn\inst{\ref{csh}, \ref{zurich}}\orcid{0000-0002-4047-4724}
         \and
         D.~J.~M.~Petit~dit~de~la~Roche\inst{\ref{geneva}, \ref{potsdam}}\orcid{0000-0002-8963-3810}
         \and
         S.~Saha\inst{\ref{chile1},\ref{chile2}}\orcid{0000-0001-8018-0264}
         \and
         A.~M.~S.~Smith\inst{\ref{dlr}}\orcid{0000-0002-2386-4341}
         \and
         S.~Ulmer-Moll\inst{\ref{leiden}}\orcid{0000-0003-2417-7006}
         \and
         P.~J.~Wheatley\inst{\ref{warwick1},\ref{warwick2}}\orcid{0000-0003-1452-2240}
          }

   \institute{
    Geneva Observatory, University of Geneva, Chemin Pegasi 51, 1290 Versoix, Switzerland\label{geneva}
    \and
    Physikalisches Institut, University of Bern, Gesellschaftsstrasse 6, 3012, Bern, Switzerland\label{bern}
    \and
    European Space Agency (ESA), European Space Research and Technology Centre (ESTEC), Keplerlaan 1, 2201 AZ Noordwijk, The Netherlands\label{estec}
    \and
    Astronomy Unit, Queen Mary University of London, Mile End Road, London E1 4NS, UK\label{queen_mary}
    \and
    Leibniz-Institut für Astrophysik Potsdam (AIP), An der Sternwarte 16, 14482 Potsdam, Germany\label{potsdam}
    \and
    Department of Physics, University of Warwick, Gibbet Hill Road, Coventry CV4 7AL, UK\label{warwick1}
    \and
    Centre for Exoplanets and Habitability, University of Warwick, Gibbet Hill Road, Coventry CV4 7AL, UK\label{warwick2}
    \and
    Department of Physics and Astronomy, Michigan State University, East Lansing, MI 48824, USA\label{michigan}
    \and
    Netherlands Institute for Radio Astronomy, ASTRON, Oude Hoogeveensedijk 4, Dwingeloo, 7991 PD, The Netherlands\label{astron}
    \and
    Institute of Physics, \'Ecole Polytechnique F\'ed\'erale de Lausanne (EPFL), Observatoire de Sauverny, Chemin Pegasi 51b, 1290 Versoix, Switzerland\label{epfl}
    \and
    Instituto de Estudios Astrofísicos, Facultad de Ingeniería y Ciencias, Universidad Diego Portales, Av. Ejército Libertador 441, Santiago, Chile\label{chile1}
    \and
    Centro de Excelencia en Astrofísica y Tecnologías Afines (CATA), Camino El Observatorio 1515, Las Condes, Santiago, Chile\label{chile2}
    \and
    Universität Potsdam, Institut für Physik und Astronomie, Karl-Liebknecht-Straße 24/25, 14476 Potsdam, Germany\label{potsdam2}
    \and
    Center for Space \& Habitability, Universität Bern, Gesellschaftsstrasse 6, 3012 Bern, Switzerland\label{csh}
    \and
    Inst. f. Teilchen- und Astrophysik, ETH Zürich, Wolfgang-Pauli-Strasse 27, 8093 Zürich, Switzerland\label{zurich}
    \and
    Institute of Space Research, German Aerospace Center (DLR), Rutherfordstr. 2, 12489 Berlin, Germany\label{dlr}
    \and
    Leiden Observatory, University of Leiden, Einsteinweg 55, 2333CA Leiden, The Netherlands\label{leiden}
    }


 
  \abstract{HIP 67522 is a one of the youngest multi-planetary systems discovered to date. The 17 Myr-old, Sun-like star is part of the Scorpius-Centarus OB association and is known to host two Saturn-sized planets in near 2:1 mean motion resonance. We analysed photometric transits observed with the CHaracterising ExOPlanet Satellite (CHEOPS), the Transiting Exoplanets Survey Satellite (TESS) and multiple ground-based facilities to search for transit timing variations induced by planet-planet gravitational interactions, as well as transit depth variations, linked with the changing coverage of active regions on the stellar surface. We do not detect any transit timing variations for HIP 67522\,b exceeding two minutes, contrary to previous results from the \emph{James Webb} Space Telescope. In addition, we found that the observed transit depth of planet\,b changes strongly over time, with a >30\% variation in amplitude. The high sensitivity and blue bandpass of the CHEOPS satellite also enabled us to identify multiple starspot crossings and measure their properties, including spot temperatures and sizes. We also measured the shear of the host star's differential rotation by modelling the out-of-transit photometric variability and concluded that HIP 67522 exhibits a supersolar differential rotation. Based on internal structure models and planet formation simulations, we find that HIP 67522\,b likely have more than 20\% of their mass made of gaseous envelopes at their current age. Evolutionary simulations further indicate that both planets in the system will experience significant photoevaporation, evolving into either Earth size planets with completely stripped atmospheres or sub-Neptune-sized planets.}



  


   \maketitle
%

\section{Introduction}
\label{sec: introduction}

HIP 67522 is one of the youngest (17$\pm$1~Myr) multi-planetary systems discovered so far, alongside AU Mic (22$\pm$3~Myr, \citealp{aumic_age}) and V1298 Tau (23$\pm$4~Myr, \citealp{v1298_tau}). The star is part of the well-characterised Scorpius-Centarus or Sco-Cen OB association, with member ages ranging from 2 to 20~Myr \citep{scocen2023}. The system hosts two confirmed planets. The inner planet b is a super-puffy Saturn-sized planet orbiting with a period of 6.96~days \citep{rizzuto2020}. Recently, the discovery of an outer sub-Saturn-sized planet with an orbital period of 14.33~days was announced \citep{barber2024}. The two planets are near a first-order mean motion resonance, with a fractional deviation from a perfect 2:1 commensurability of 2.9\% \citep{dai2024}. As only a handful of young multi-planetary systems are known, HIP 67522 has been extensively studied through both ground- and space-based observations. 

\citet{heitzmann2021}, using ground-based transit spectroscopy, measured the projected spin-orbit obliquity angle of HIP 67522\,b to be |$\lambda$| = 5.8$^{+2.8\circ}_{-5.7}$, and the derived 3D obliquity to be $\psi$ = 20.2$^{+10.3\circ}_{-8.7}$. This small angle indicates that the planet probably did not form through high-eccentricity migration. Doppler tomography revealed the presence of dark spots in two separate transit observations. This, along with the high (5\%) amplitude flux variability in the TESS light curves, indicates that the host star is highly active.

\citet{Thao2024} presented the \emph{JWST} NIRSpec/G395H transmission spectrum of HIP 67522 b. The spectrum revealed high (30-50\%) amplitude features due to the presence of H$_{2}$O, CO$_{2}$ and CO. Using the amplitude of these features, \citet{Thao2024} placed an upper limit of $<$ 20M$_{\oplus}$ on the planetary mass. The atmospheric metallicity is constrained to be supersolar (3 to 10 $\times$ solar), and the range of possible C/O ratio ranges from solar to subsolar. In addition, the NIRSpec white-light curve also reveals two separate spot-occultation events.  

Utilising XUV observations, \citet{maggio2024} derived that the host star has a very hot corona with plasma temperatures reaching 20 million Kelvin, even in the quiescent state. Additionally, 12 flare events were identified in the three TESS sectors and, \citet{Ilin2025} presented a correlation between the clustering of flaring events and the orbital phase of the inner planet b, which points towards the presence of star-planet interactions in the system.

One of the main challenges of characterising planets around young stars is mitigating the effect of stellar activity. Both occulted and unocculted stellar spots or faculae can cause significant biases in the measured transit depth and, consequently, in the density measurement. It can also alter the measured transmission spectrum through the transit light source effect or ``stellar contamination" \citep{pont2008,rackham2018,sage}. However, planets around active stars provide a unique opportunity to map the activity features on the stellar surface and study their properties \citep{silva2003, desert2011,sanchis2011,wasp-85a,morris2017, chakraborty2025}. 


Differential rotation is an important factor affecting the evolution of active regions. The strength or shear of differential rotation is stronger for young stars than for their adult counterparts \citep{McQuillan2014}. In general, several techniques are available to detect the Differential Rotation (DR) of a star including the Reloaded Rossiter-McLaughlin effect \citep{Cegla2016,doyle2023}, Doppler Imaging \citep{vogt1983}, the Fourier transform method \citep{reiners2003}, and time series photometry. Using the Kepler light curves, \citealp{reinhold2015} presented a technique for measuring DR by analyzing the presence of multiple periodicities in the Lomb-Scargle periodogram of the light curve. While the technique is sensitive to detecting DR, the periodicities of the secondary peaks vary depending on the method used to calculate them, which influences the measurement of absolute shear ($\Delta\Omega$) and the rotation period ($\alpha$). 

The CHaracterising ExOPlanet Satellite (CHEOPS; \citealp{benz_cheops}) has been used to study transiting planets around young stars such as AU Mic \citep{aumic_paper1, aumic_paper2} and V1298 Tau \citep{Damasso2023}. By measuring precise transit times, CHEOPS has been used to constrain planetary masses and system architectures through Transit Timing Variations (TTVs). It has also been used to search for activity-induced Transit Depth Variations (TDeVs) in young systems. In this work, we extend these efforts by using CHEOPS to characterise the keystone HIP 67522 system.

The paper is structured as follows: in Section \ref{sec: observations}, we describe our CHEOPS, TESS and ground-based follow-up observations. In Section \ref{sec: planetary_properties}, we present our transit analysis, describing our fitting technique. In Section \ref{sec: stellar_properties}, we determine the properties of the host star. In Section \ref{sec: discus}, we discuss our results and in Section \ref{sec:concl}, we provide our conclusions.

\section{Observations}
\label{sec: observations}

\subsection{CHEOPS}
\label{sec:cheops_sec}

We observed four full transits and 11 partial transits of HIP 67522 b with CHEOPS \citep{benz_cheops} between March and June 2024. Six additional out-of-transit observations were scheduled to aid the transit observations by characterising the evolving activity of the host star. The observations were conducted as part of two different GO programs: PR240004 (PI: Ilin) and PR240017 (PI: Chakraborty). The log of our observations is provided in Table \ref{tab: CHEOPS_logs}. A Gaia G band magnitude of 9.613 enabled high-cadence photometry with an exposure time of 10 seconds. To limit the downlink data rate, CHEOPS is designed to stack science images onboard below a certain exposure time. Here, every three consecutive images are co-added, resulting in an effective exposure time of 30 seconds. Smaller imagettes at the requested exposure time (10 seconds) are also downlinked. The photometric light curve is automatically extracted from the science images using the CHEOPS Data Reduction Pipeline (DRP, \citealp{drp}). However, here we relied on the PSF Imagette Photometric Extraction (PIPE, \citealp{pipe}) pipeline to recover the light curves. PIPE performs PSF fitting (while the DRP relies on aperture photometry) directly on the imagettes, yielding a higher cadence light-curve. PIPE is also more effective at mitigating photometric variations induced by background stars and modulated by the spacecraft roll-angle. On average, only 62.6\% of each visit contains data. We clipped all identifiable flares from our data before proceeding with our analysis. The timestamps of the clipped flares are listed in Extended Data Table 3 of \cite{Ilin2025}.

\subsection{TESS}
For our analysis, we used all available TESS \citep{ricker_tess} observations of HIP 67522, which include sectors 11, 38 and 64. The mid-sector times are May 2019, May 2021 and May 2023, respectively. Each sector contains four transits of planet b. These transits are denoted as S\{sector number\}.\{transit number\} in this paper. The data were downloaded from the Mikulski Archive for Space Telescopes (MAST) using the \texttt{lightkurve} package \citep{lightkurve}. We choose to use the light curves obtained from the Presearch Data Conditioning Simple Aperture Photometry (PDCSAP) pipeline, as they are corrected for instrumental systematics by the TESS Science Processing Operation Center (TESS-SPOC, \citealp{tess_spoc}). 

For sector 11, two transits (S11.1 and S11.3) are not available as a part of the PDCSAP data. Thus, we choose to use the Simple Aperture Photometry (SAP) light curves for these transits.

\label{sec:tess_sec}

\subsection{EulerCam \& NGTS}
\label{sec:ground_sec}

In parallel with our CHEOPS observations, we monitored the flux variability of HIP 67522 using EulerCam \citep{Lendl2012}. EulerCam is a 4k$\times$4k CCD camera installed in the Cassegrain focus of the 1.2-meter Euler telescope at ESO's La Silla observatory. We monitored the flux variability on a nightly basis from April 1, 2024 to June 30 2024 in four different filters: Geneva-$B$, Geneva-$V$, Gunn-$r'$, and Gunn-$z'$. We set the exposure times to 60, 30, 20 and 15 seconds, respectively.

\begin{figure}
   \centering
    \resizebox{\columnwidth}{!}{\includegraphics[trim=0.0cm 0.0cm -3.0cm 0.0cm]{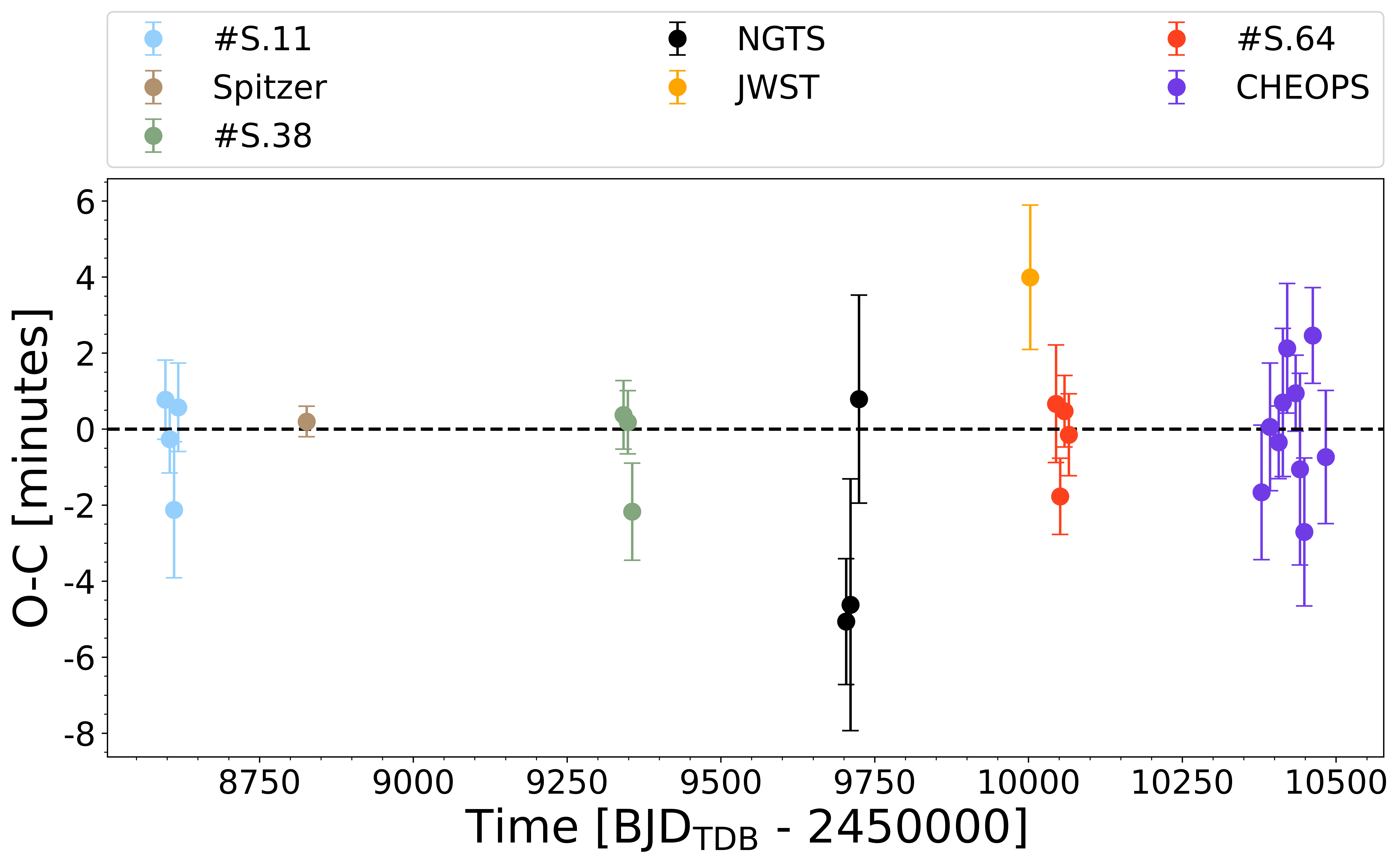}}
      \caption{Difference between the observed and calculated mid-transit times, based on a linear ephemeris, for individual transits of HIP 67522\,b across different epochs. The transit timings for the Spitzer and JWST observations are taken from \citet{rizzuto2020} and \citet{Thao2024}, respectively. }
         \label{fig: ttvs}
\end{figure}

The acquired raw images were corrected for bias, flat and overscan \citep{Lendl2012}, before performing aperture photometry. We used circular apertures for the target and reference stars with radii ranging from 10 to 80 pixels. Due to pointing inaccuracies, the exact location of stars on the detector varied slightly from night to night. We customised the standard EulerCam reduction pipeline to position apertures on the right ascension and declination of the stars rather than their X-Y location on the detector. For each image, we obtained the necessary astrometric solution using the \texttt{astrometry.net} package \citep{astrometry_net}. The optimal aperture for each night was determined by minimizing the RMS scatter for the nightly images. 

The light curves show clear signs of stellar activity-induced photometric variability at the star's rotation period. The amplitude of the variability depends on the bandpass. Additionally, we observed part of the high-amplitude CHEOPS flare that occurred on 2024-03-23 (file\_key: CH\_PR240004\_TG000103\_V0300) in all four filters. This flare is masked and excluded from further analysis in this paper. 

We also utilised 92 nights of photometric monitoring by the Next Generation Transit Survey (NGTS; \citealp{wheatley2018}). NGTS consists of 12 fully automated 20-cm telescopes located at the ESO's Paranal observatory, Chile. Each NGTS camera observes a wide (2.8$^{\circ}$$\times$2.8$^{\circ}$) field of view with 13 second cadence using a custom 520-890 nm filter. HIP 67522 was mostly observed in blind-survey mode between May 2022 and September 2023, wherever possible using the same individual telescope and similar air-masses to allow for consistent guiding and comparable night-by-night photometry. Observation durations varied between 30 minutes and the entire observing night, depending on nightly target visibility and the wider availability of the NGTS telescopes. Additionally, three transits of planet b were acquired using two telescopes on the nights of 2022-05-04, 2022-05-10 and 2023-06-02 to aid the wider system characterisation. The light curves were extracted using the standard NGTS aperture photometry pipeline, and the systematics removal was performed using the technique detailed in \citet{wheatley2018}.

\begin{figure}
   \centering
    \resizebox{\columnwidth}{!}{\includegraphics[trim=0.0cm 0.0cm -3.0cm 0.0cm]{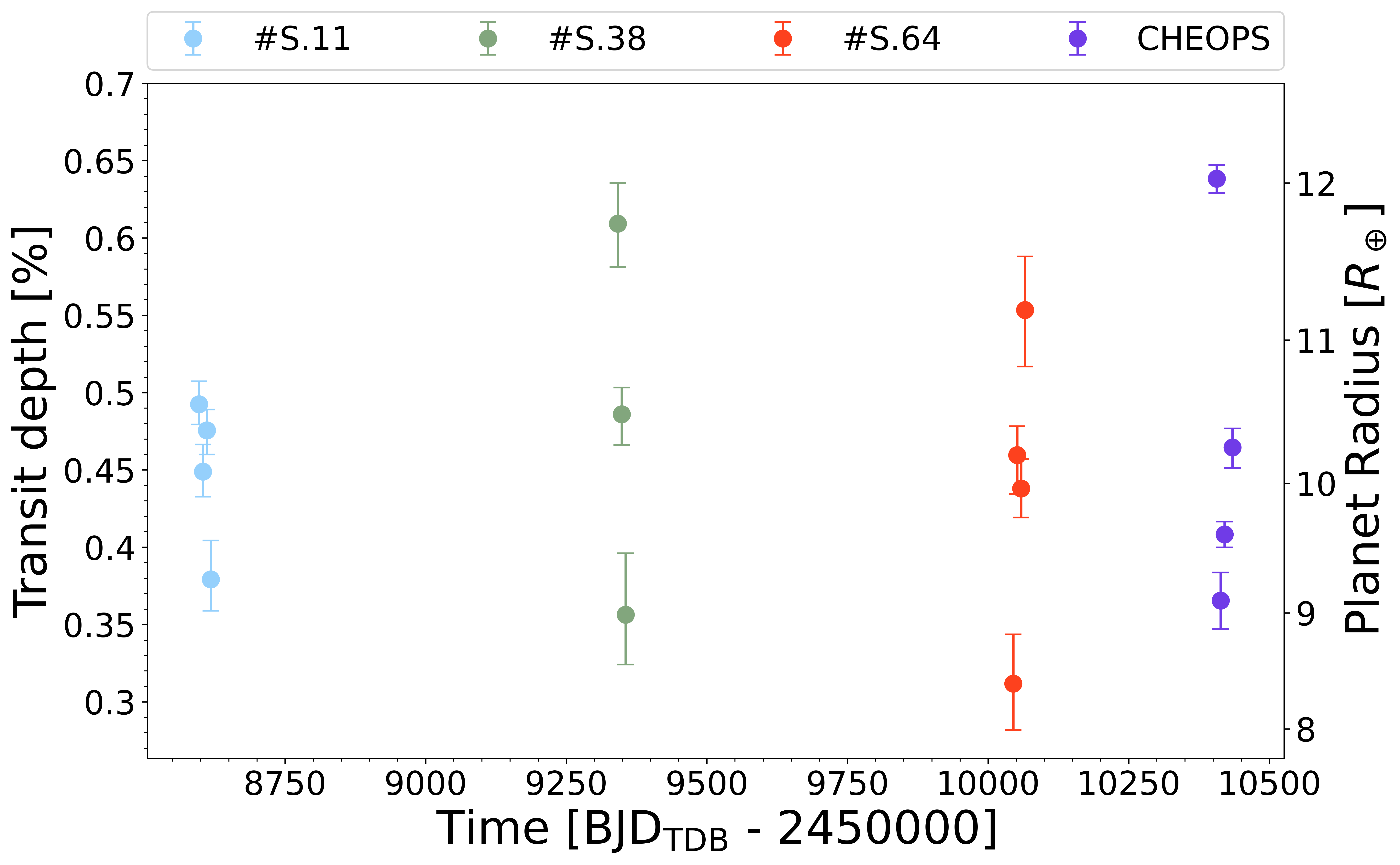}}
      \caption{Observed transit depth variations for HIP 67522\,b using CHEOPS and TESS observations. The secondary axis displays the derived planetary radii assuming a stellar radius of 1.38$\pm$0.06 R$_{\odot}$}
         \label{fig: tdevs}
\end{figure}

\section{The HIP 67522 planetary system}
\label{sec: planetary_properties}

\subsection{Revised planet parameters}
\label{subsec: sys_params}

To determine the system parameters, we simultaneously fitted all light curves using the COde for exoplaNet ANalysis (\texttt{CONAN}\footnote{\url{https://github.com/titans-ge/CONAN}}, \citealp{akinsanmi2025,lendl2020}). We set wide Gaussian priors on the mid-transit time (T$_{0}$) and orbital period (P$_{\rm{orb}}$), centered around estimates from \citet{barber2024}. For stellar density ($\rho_{\star}$), planet-to-star radius ratio (R$_{\rm{pl}}$/R$_{\star}$) and impact parameter (b), we set wide uniform priors. Additionally, we fixed the eccentricity and argument of periastron to 0 and 90$^{\circ}$, respectively, following previous studies \citep{heitzmann2021, Thao2024}. To model limb-darkening, we used the quadratic limb-darkening law \citep{Kopal1950}. We set normal priors on the two coefficients centered around estimates obtained from the limb-darkening and uncertainties package (\texttt{LDCU}\footnote{\texttt{LDCU} is a modified version of the python routine implemented by \citet{espinoza2015} that computes the limb-darkening coefficients and their corresponding uncertainties using a set of stellar intensity profiles accounting for the uncertainties on the stellar parameters. The stellar intensity profiles are generated based on two libraries of synthetic stellar spectra: ATLAS \citep{kurucz1979} and PHOENIX \citep{husser2013}.}). In addition, we impose the boundary conditions by \citet{kipping2013}.

Apart from the transit signals, we must account for the rapid out-of-transit flux variability of the star. To model the stellar variability and instrumental systematics simultaneously, we used a mixture of parameter models and Gaussian Process (GP) regression. The CHEOPS light curves display well-known correlations with background flux and spacecraft roll angle \citep{lendl2020,akin2024}. We fit a baseline model with a second-order polynomial on both the background and the cosine of the roll angle. For each filter, the GP model is calculated using \texttt{celerite} \citep{celerite}, and we used the Simple Harmonic Oscillator (SHO) kernel over time, which is commonly used to model rotationally modulated flux variability \citep{aumic_paper1, aumic_paper2}. The free kernel hyper-parameters include the amplitude ($\sigma$),  period (P$_{0}$) and quality factor ($Q$). We set wide log-uniform priors on the amplitude ranging from 0 to 10$^{6}$ ppm, period from 0 to 3 days, and quality factor from 0 to 100. A complete list of our priors, fitted and derived parameters is presented in Table.~\ref{table:orbital_solution}. The transit light curves with the constrained best-fit models is shown in Fig.~\ref{fig:cheops_ngts_tess_lightcurves}.

\begin{table*}
\caption{Fitted and derived parameters for the planets presented in this paper.}\label{table:orbital_solution}

\centering
\begin{tabular}{l c c c}
\toprule
\noalign{\smallskip}
Parameter, Symbol, Units & Priors & HIP 67522 b & HIP 67522 c \\ 
\hline
\noalign{\smallskip}
\multicolumn{4}{c}{Fitted Planetary Parameters} \\
\noalign{\smallskip}
\hline
\noalign{\smallskip}
Mid-Transit Time, $T_0$ [BJD$_\mathrm{TDB}$] & $\mathcal{N}(2458604.0, 0.1)$& $2458604.0243\pm0.0003$ & $2458602.5029\pm0.0008$ \\
 & $\mathcal{N}(2458602.5, 0.1)$&  &  \\
\noalign{\smallskip}
Orbital Period, $P_{\rm{orb}}$ [days]  & $\mathcal{N}(6.9, 0.1)$ & $6.959469\pm0.000002$ & $14.334886\pm0.000011$ \\
  & $\mathcal{N}(14.3, 0.1)$ &  &  \\
\noalign{\smallskip}
Radius Ratio, $R_\mathrm{pl}/R_{\star}$ & $\mathcal{U}(0.0, 0.1)$ & $0.056 - 0.0799$$^{\dagger}$ & $0.0512\pm0.0016$ \\
\noalign{\smallskip}
Impact Parameter, $b$ & $\mathcal{U}(0, 1)$ & $0.195_{-0.066}^{+0.069}$  & $0.432_{-0.024}^{+0.030}$   \\
\noalign{\smallskip}
Eccentricity, $e$ & & 0& 0\\
\noalign{\smallskip}
Argument of Periastron, $\omega$ [$^{\circ}$] & & 90& 90\\
\noalign{\smallskip}
\hline
\noalign{\smallskip}
\multicolumn{4}{c}{Fitted Stellar Parameters} \\
\noalign{\smallskip}
\hline
\noalign{\smallskip}
Stellar Density, $\rho_{\star}$ [g cm$^{-3}$]  & $\mathcal{U}(0.2, 1.0)$ & \multicolumn{2}{c}{$0.585_{-0.028}^{+0.021}$}    \\
\noalign{\smallskip}
LD Coefficient (CHEOPS), $q1_{\rm{CHEOPS}}$ & $\mathcal{N}(0.46, 0.01)$ & \multicolumn{2}{c}{$0.45\pm0.01$} \\
\noalign{\smallskip}
LD Coefficient (CHEOPS), $q2_{\rm{CHEOPS}}$ & $\mathcal{N}(0.32, 0.02)$ & \multicolumn{2}{c}{$0.32\pm0.02$} \\
\noalign{\smallskip}
LD Coefficient (TESS), $q1_{\rm{TESS}}$ & $\mathcal{N}(0.34, 0.01)$ & \multicolumn{2}{c}{$0.34\pm0.01$} \\
\noalign{\smallskip}
LD Coefficient (TESS), $q2_{\rm{TESS}}$ & $\mathcal{N}(0.27, 0.02)$ & \multicolumn{2}{c}{$0.27\pm0.02$} \\
\noalign{\smallskip}
LD Coefficient (NGTS), $q1_{\rm{NGTS}}$ & $\mathcal{N}(0.40, 0.01)$ & \multicolumn{2}{c}{$0.39\pm0.01$} \\
\noalign{\smallskip}
LD Coefficient (NGTS), $q2_{\rm{NGTS}}$ & $\mathcal{N}(0.29, 0.02)$ & \multicolumn{2}{c}{$0.29\pm0.02$} \\
\noalign{\smallskip}
GP Amplitude (CHEOPS), $\sigma_{\rm{CHEOPS}}$ [ppm] & $\mathcal{LU}(0.007, 10^{6}$) &\multicolumn{2}{c}{$13507_{-3798}^{+4590}$}\\
\noalign{\smallskip}
GP Timescale (CHEOPS), $P_{0, \rm{CHEOPS}}$ [days] & $\mathcal{LU}(0.007, 3)$ & \multicolumn{2}{c}{$0.104_{-0.042}^{+0.062}$} \\
\noalign{\smallskip}
GP Quality (CHEOPS), $Q_{\rm{CHEOPS}}$ & $\mathcal{LU}(0, 10^{2})$ & \multicolumn{2}{c}{$0.0030_{-0.0008}^{+0.0011}$} \\
\noalign{\smallskip}
GP Amplitude (TESS), $\sigma_{\rm{TESS}}$ [ppm] & $\mathcal{LU}(0.007, 10^{6}$) &\multicolumn{2}{c}{$14107_{-910}^{+1094}$}\\
\noalign{\smallskip}
GP Timescale (TESS), $P_{0, \rm{TESS}}$ [days] & $\mathcal{LU}(0.007, 3)$ & \multicolumn{2}{c}{$1.24\pm0.02$} \\
\noalign{\smallskip}
GP Quality (TESS), $Q_{\rm{TESS}}$ & $\mathcal{LU}(0, 10^{2})$ & \multicolumn{2}{c}{$4.11_{-0.50}^{+0.63}$} \\
\noalign{\smallskip}
GP Amplitude (NGTS), $\sigma_{\rm{NGTS}}$ [ppm] & $\mathcal{LU}(0.007, 10^{6}$) &\multicolumn{2}{c}{$19041_{-5777}^{+11215}$}\\
\noalign{\smallskip}
GP Timescale (NGTS), $P_{0, \rm{NGTS}}$ [days] & $\mathcal{LU}(0.007, 3)$ & \multicolumn{2}{c}{$0.86_{-0.27}^{+0.54}$} \\
\noalign{\smallskip}
GP Quality (NGTS), $Q_{\rm{NGTS}}$ & $\mathcal{LU}(0, 10^{2})$ & \multicolumn{2}{c}{$0.43_{-0.23}^{+0.61}$} \\
\noalign{\smallskip}

\noalign{\smallskip}
\hline
\noalign{\smallskip}
\multicolumn{4}{c}{Derived parameters} \\
\noalign{\smallskip}
\hline
\noalign{\smallskip}
\multicolumn{2}{l}{Planetary Radius, $R_\mathrm{pl}$ [$R_\mathrm{Jup}$]}  & $0.75 - 1.07$$^{\dagger}$  & $0.69\pm 0.04$  \\
\noalign{\smallskip}
\multicolumn{2}{l}{Planetary Radius, $R_\mathrm{pl}$ [$R_{\oplus}$]}  & $8.4 - 12.0$$^{\dagger}$  & $7.69\pm 0.44$  \\
\noalign{\smallskip}
\multicolumn{2}{l}{Inclination, $i$ [$^{\circ}$]}  & $89.02_{-0.37}^{+0.34}$  & $88.66_{-0.11}^{+0.09}$  \\
\noalign{\smallskip}
\multicolumn{2}{l}{Scaled Semi-Major Axis, $a_\mathrm{pl}/R_\star$} & $11.44_{-0.18}^{+0.13}$  & $18.5_{-0.3}^{+0.2}$  \\
\noalign{\smallskip}
\multicolumn{2}{l}{Semi-Major Axis, $a$ [AU]} & 0.073$\pm$0.003 & 0.119$\pm$0.005 \\
\noalign{\smallskip}
\multicolumn{2}{l}{Transit Duration, $T_{14}$ [hours]} & $4.87\pm 0.01$  & $5.66\pm 0.02$  \\
\noalign{\smallskip}
\multicolumn{2}{l}{Transit Depth, $\delta$ [\%]} & 0.44$\pm$0.01  & 0.26$\pm$0.02  \\
\noalign{\smallskip}
\multicolumn{2}{l}{Planetary Mass, $M_{\rm{pl}}$ [$M_{\rm{\oplus}}$]} & 13.8$\pm$1.0$^{\ddagger}$ & $\leq$22$^{\ddagger}$  \\
\noalign{\smallskip}
\bottomrule
\noalign{\smallskip}
\multicolumn{4}{l}{
\begin{minipage}{0.9\linewidth}
\small
\textit{Notes:} $\dagger$ Maximum and minimum radius values from the Section 3.2 analysis. \\ $\ddagger$ The mass is constrained from dynamical modelling of transit timing variations assuming non co-planarity with the inner planet b. The mass of planet b is taken from \cite{Thao2024}.
\end{minipage}
}
\end{tabular}
\end{table*}

\subsection{Transit timing and depth variations}
\label{subsec: ttvs_and_tdevs}

The inner planet b is near resonance with the outer planet c. The fractional deviation from perfect 2:1 period commensurability is $\sim$3\%. To search for any timing variations associated with the resonance, we fit the mid-transit times for all transits acquired from TESS and CHEOPS individually. To keep the model fitting tractable, we clipped the TESS transit light curves using a 10-hour window around the transits. For CHEOPS and TESS light curves, we fit the stellar variability with a fourth- and sixth-order polynomial for partial and full transits, respectively. The remaining correlated noise, arising from  unaccounted instrumental systematics or stellar phenomena such as micro-flares, and active-region occultations, is accounted for using a GP with real term kernel implemented in \texttt{celerite}. The choice of the kernel and the order of the polynomial is motivated by model comparison, maximising the Bayesian evidence, to avoid over-fitting. The shorter duration, which is approximately half of a full transit observation, also justifies a lower-order polynomial for partial transits. We first fit the polynomial to the out-of-transit flux and then fix the coefficients to the derived values. The GP kernel parameters, amplitude and period, are set as free parameters with log-uniform priors ranging from 0 to 10$^{3}$ ppm and from 0 to 1 day, respectively. For CHEOPS light curves, we also fit a baseline model made of a second-order polynomial on the background and the cosine of the roll angle. For the transit model, only the mid-transit time and transit depth are set as free parameters, and the rest are fixed to the values derived in Sec. \ref{subsec: sys_params}.  

In addition, we used the NGTS light curves for our timing analysis. We fitted the individual light curves setting a broad uniform prior on the mid-transit times, transit depth, and SHO GP hyperparameters. The measured mid-transit times for individual transits are listed in the Table \ref{tab: transit_times_radius_ratio}. The CHEOPS partial transit observations with no visible ingress/egress are excluded from our timing variation analysis, as they are less precise with typical uncertainties $\geq$ 5 min.

With individual mid-transit times, we performed an O-C analysis similar to \citet{sterken2005}, calculating the deviation of the observed mid-transit times from the predictions from a mono-periodic Keplerian model: C = T$_{0}$ + E $\times$ P. Here, E is the epoch transit number, T$_{0}$ is the reference mid-transit time, and P is the orbital period of the planet. We jointly fit for the reference mid-transit time and orbital period. The result of our analysis is shown in Fig. \ref{fig: ttvs}.

\begin{figure*}
   \centering
\resizebox{\textwidth}{!}{\includegraphics[trim=0.0cm 0.0cm 0.0cm 0.0cm]{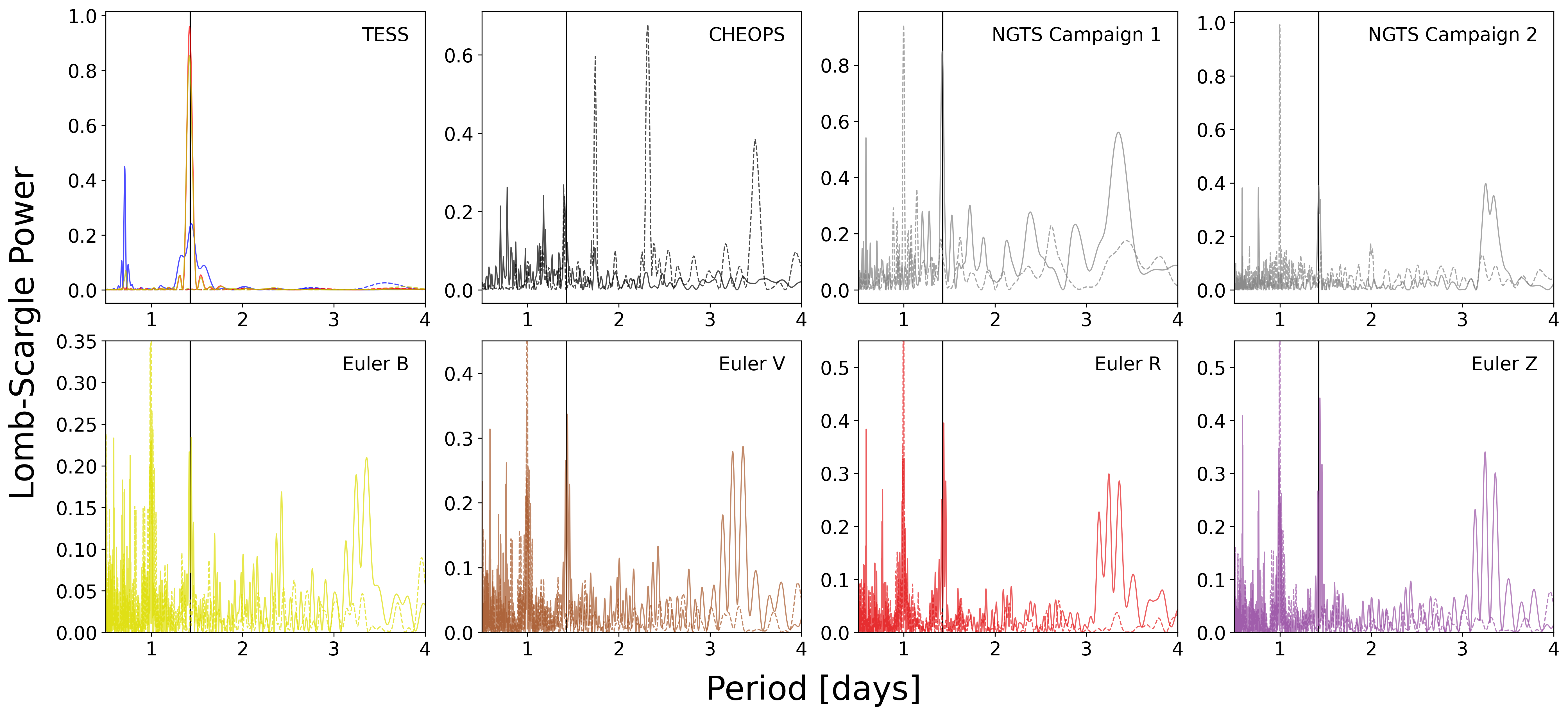}}
      \caption{Generalised Lomb-Scargle periodogram computed for different light curves. The TESS, NGTS and multi-band Euler light curves show clear evidence of stellar rotation-induced periodicity at 1.42 days. The CHEOPS periodogram is limited by its complex observing window. The measured rotation period is marked in solid black, and the dashed lines represent the corresponding window functions.}
         \label{fig: gls_periodograms}
\end{figure*}

\begin{figure}
   \centering
\resizebox{\columnwidth}{!}{\includegraphics[trim=0.0cm 0.0cm -3.0cm 0.0cm]{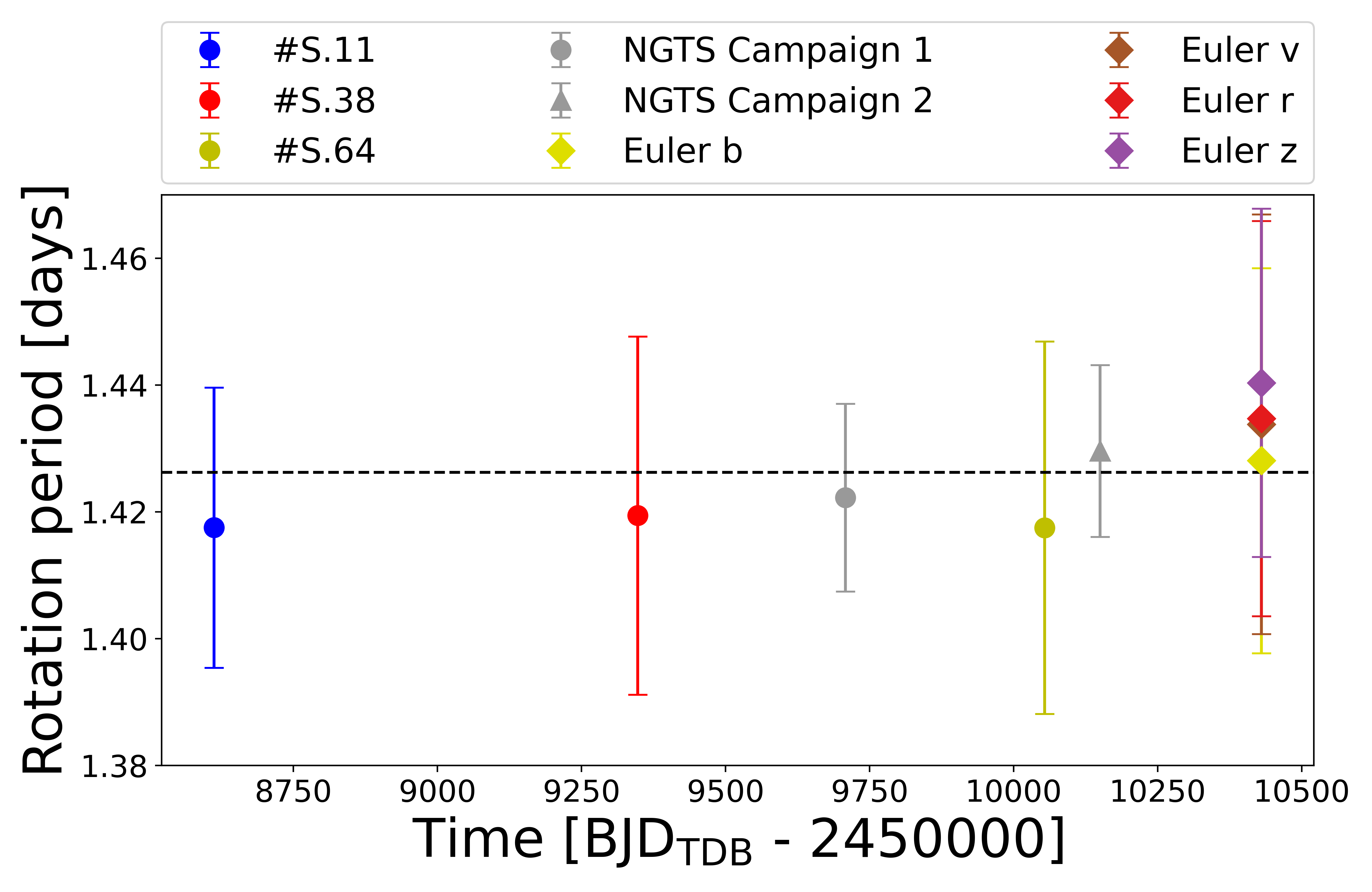}}
      \caption{ Rotation period of HIP 67522 across different observation cycles. The dashed line indicates the error-weighted average of all measurements. Overall, the inferred stellar rotation period remains stable over time. In the TESS panel, sectors 11, 38 and 64 are shown as blue, red and yellow solid lines, respectively.}
         \label{fig: rotation rate evolution}
\end{figure}

The presence of stellar activity impacts the measurement of the transit depth. If active regions are present directly on the transit chord, the occultations of the active regions by the planet result in light curve distortions. In the case of unocculted active regions, the observed transit depth can be shallower or deeper than the true transit depth, depending on the presence of faculae or spots, respectively (see e.g., \citealp{Moran2023,dominique2024}). The amplitude of these effects scales with the contrast and filling factor of active regions. We expect the impact to be higher in the CHEOPS than in the TESS data, due to its bluer bandpass.

To detect any transit depth variations, we measured the transit depth for individual TESS and CHEOPS observations. Since both TESS and CHEOPS probe at optical wavelengths, the measured transit depths can be directly compared. We excluded all CHEOPS partial transits, and NGTS observations from our transit depth variation analysis, due to the lack of a pre-transit baseline, and differences in filter bandpasses. The TESS S38.1 transit is also excluded, as the observation is strongly affected by flares. 

As illustrated in Fig.~\ref{fig: tdevs}, the TESS and CHEOPS transits show high-amplitude transit depth variability, with a peak-to-peak variability amplitude of $>$30\%. This variability is far greater than the measurement uncertainty and is evident even between consecutive transits, suggesting a highly active stellar surface. Such high-amplitude TDeV is extremely rare; the most comparable case is AU Mic, which shows an amplitude of $\sim$20\% \citep{aumic_paper1}. This makes HIP 67522 one of the most compelling targets for studying the influence of stellar surface inhomogeneities on exoplanet transits.

\section{Host star activity and differential rotation}
\label{sec: stellar_properties}

\subsection{Stellar rotation period determination}
\label{subsec: rotation_period}

The TESS, NGTS, Euler and CHEOPS light curves show rapid out-of-transit flux variability due to the presence of stellar active regions and stellar rotation. These light curves are obtained over several seasons and allow us to measure the variability in the measured rotation rate arising from the evolution of active regions and the differential rotation of the star. For this, we computed the Generalised Lomb-Scargle (GLS) periodogram for each light curve \citep{lomb, scargle, astropy}. We also computed the window function for each observation to identify the peak periodicities associated with aliasing.

From three TESS sectors, we have near continuous monitoring of the stellar variability for 26 days, each. Sectors 38 and 64 show significant peaks at roughly 1.42 days. This is in line with the rotation rate calculated by \citet{rizzuto2020}. For TESS sector 11, the highest peak is found at 1.42/2=0.71 days, that is, half the rotation period. This is caused by the double-dip profile of the light curve in the first half of the observations (Fig. \ref{fig: sage_model}).

For the CHEOPS light curve, the complex window function limits us from constraining the rotation period (Fig. \ref{fig: gls_periodograms}). Hence, we omit them from our rotation period determination. 

For NGTS, we divided the light curve into two separate datasets. The observations obtained between 2022-04-16 and 2022-05-28 were grouped together as a set (campaign 1), and the data acquired between 2023-06-01 and 2023-08-22 were grouped as a second set (campaign 2).  This division was done to avoid equally spaced aliases around the stellar rotation period, which could arise due to the $\sim$1-year gap between the two NGTS campaigns (e.g. \citealp{aumic_paper1}). The periodograms obtained are shown in Fig. \ref{fig: gls_periodograms}. A strong peak is observed around the expected rotation rate, along with other peaks arising from the aliasing effect. The most prominent aliases within our frequency search grid were calculated using the methods outlined in \citet{vanderplas2018}, to check any overlap with the measured stellar rotation period. 

For the multiband Euler light curves, we calculated the periodogram individually for each filter. A clear periodicity at the rotation rate of the star is visible in all light curves, along with the most prominent aliases. The peak periodicity is split into multicomponents, which can be interpreted as surface differential rotation of the star \citep{reinhold2015}, see Sec. \ref{subsec: evolution_of_active_regions}.

The measured rotation periods are shown in Fig. \ref{fig: rotation rate evolution}. For TESS sector 11, we report double the peak periodicity to account for the light curve morphology. We calculated the rotation period for each observation by fitting a Gaussian to the peak periodicity, and the error-weighted rotation period of the star is 1.4262 $\pm$ 0.0073 days. The rotation period and the orbital periods of planets\,b and c exhibit near spin-orbit commensurabilities of 5:1 and 10:1, with fractional deviations\footnote{The fractional deviation is defined as $\Delta$ = $\frac{P_{\rm{spin}}/P_{orb}}{p/q} - 1$, where p and q are integers that define the commensurability.} of $2.4\%$ and 0.5\%, respectively.

\subsection{Spot occultations during transit}
\label{subsec: spot_occultations}

\begin{figure}
   \centering
\resizebox{\columnwidth}{!}{\includegraphics[trim=0.0cm 0.0cm -0.0cm 0.0cm]{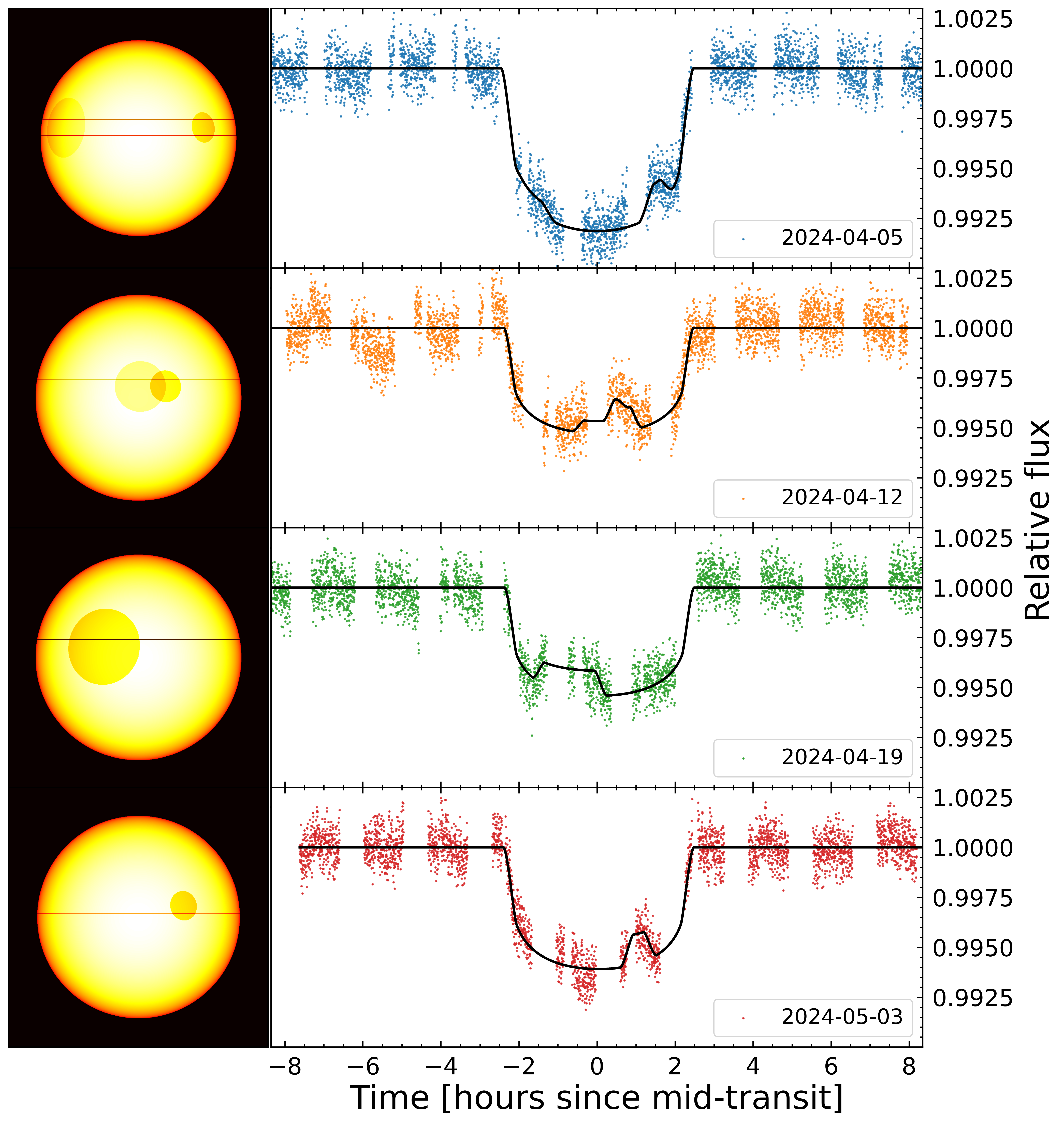}}
      \caption{CHEOPS light curves of HIP 67522 b. Left panels: Projected stellar disc showing the transit chord, limb-darkening, and starspots. The representation is to-scale. Right panels: De-trended light curves with the best-fit spot-transit model. The transit depth variations associated with unocculted active regions are clearly visible.}
         \label{fig: cheops_spot_occultations}
\end{figure}

\begin{figure}
   \centering
\resizebox{\columnwidth}{!}{\includegraphics[trim=0.0cm 0.0cm -0.0cm 0.0cm]{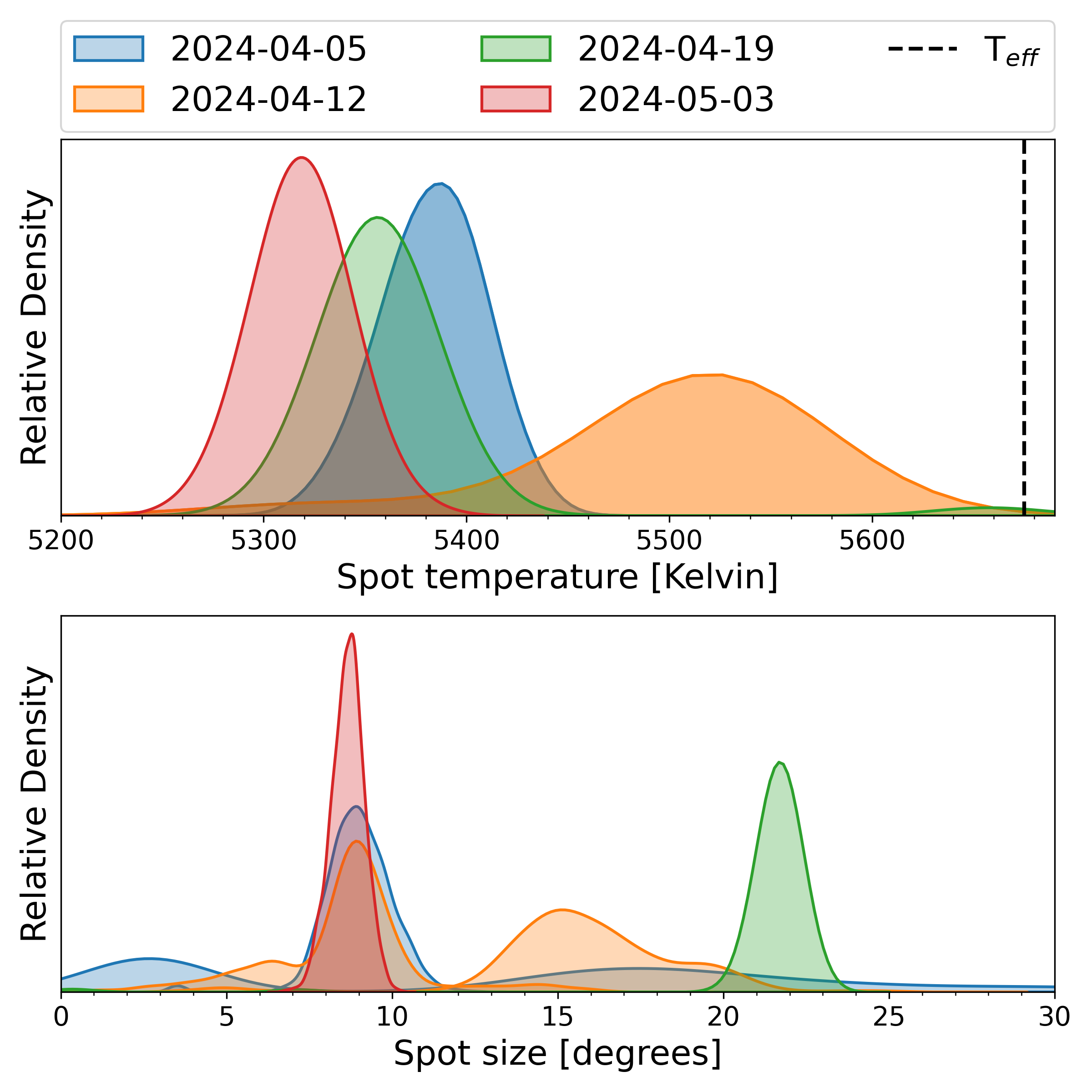}}
      \caption{Angular size and temperature of starspot occultations detected in the CHEOPS light curves of HIP 67522 b. The associated light curves are shown in Fig. \ref{fig: cheops_spot_occultations}.}
         \label{fig: cheops_spot_temp}
   \end{figure}

To search for any active region occultations by the inner planet b, we fitted the TESS and CHEOPS light curves with \texttt{PyTranSpot} \citep{pts2018}. This tool uses a pixellation approach that projects a stellar sphere and a transiting planet on a 2-dimensional Cartesian grid, whilst simulating the presence of active regions on the transit chord. The parameters for the light curve model are mid-transit time, planet-to-star radius ratio, orbital inclination, semi-major axis, orbital period, eccentricity, argument of periastron, and two (quadratic) limb darkening coefficients. An active region is parameterised by their latitude, longitude, size, and contrast with respect to the unspotted stellar photosphere. 

To limit the number of free parameters, we fixed all planetary parameters to the values derived in Sec.~\ref{subsec: sys_params}, except for the mid-transit time and the planet-to-star radius ratio. The stellar variability is modeled with a sixth-order polynomial. For CHEOPS light curves, we accounted for the background flux and the roll-angle systematics by jointly fitting two second-order polynomials in these parameters to the data. We set wide uniform priors on the longitude, size, and contrast of the active regions. The latitude of the spot was fixed to 83.5$^{\circ}$, the center of the transit chord. In addition, the stellar inclination is fixed to 90$^{\circ}$, following \cite{heitzmann2021}. To aid in convergence, we used initial guesses for the spot longitudes that we derived  using a technique similar to \citet{morris2017}, fitting a pure transit model to the light curve and convolving the residuals with a Gaussian kernel. If no prominent peaks were identified, we used a random longitude between -85$^{\circ}$ and 85$^{\circ}$ as an initial guess.

We set uniform priors on the spot longitude to cover the entire stellar surface $\mathcal{U}(-90^{\circ}, 90^{\circ})$. The uniform priors on the spot size range from small Sun-like spots (2$^{\circ}$) to large spots similar to those on M-dwarfs (31$^{\circ}$, \citealp{chakraborty2025}), and the uniform priors in the contrast range from 0 to 2, where a value $<$ 1 signifies spots, and $>$ 1 signifies faculae. To determine the number of active region occultations, we iteratively increased the number of active regions from zero to three and then compared the Bayesian Information Criterion (BIC) of the resulting fits. The model with the lowest BIC is chosen as the best-fit. 

We detected multiple spot occultations in both TESS and CHEOPS light curves (Fig.~\ref{fig: cheops_spot_occultations} \& \ref{fig:photLightcurve}). Assuming that spots and stellar photospheres both emit as blackbodies, we converted the measured contrast of active regions to temperatures \citep{silva2003}. The spot sizes and temperatures derived from the CHEOPS light curves are shown in Fig. \ref{fig: cheops_spot_temp}. The spot temperatures range from 5300 to 5600 K, approximately 500 K cooler than the clear stellar photosphere, similar to spots on the Sun \citep{Solanki2003}. The TESS light curves have a lower signal-to-noise ratio. However, we were able to detect and constrain the spot properties for a few spot occultations at S11.4, S64.2, S64.4, etc. The measured spot sizes and temperature are similar to those detected with CHEOPS (Fig. \ref{fig: tess_spot_temp}). 

Furthermore, we find no evidence for facula occultations. This is in line with Doppler tomography results from \citet{heitzmann2021}, who detected unocculted spots on the stellar surface for two separate transits, but no faculae. In addition, the JWST NIRSpec transit white light curve also showed two spot occultations \citep{Thao2024}. All these observations, including ours, were taken over multiple years, indicating that HIP 67522 is a spot-dominated star. 

\subsection{Evolution of active regions}\label{subsec: evolution_of_active_regions}

\begin{figure*}
   \centering
\resizebox{\textwidth}{!}{\includegraphics[trim=0.0cm 0.0cm 0.0cm 0.0cm]{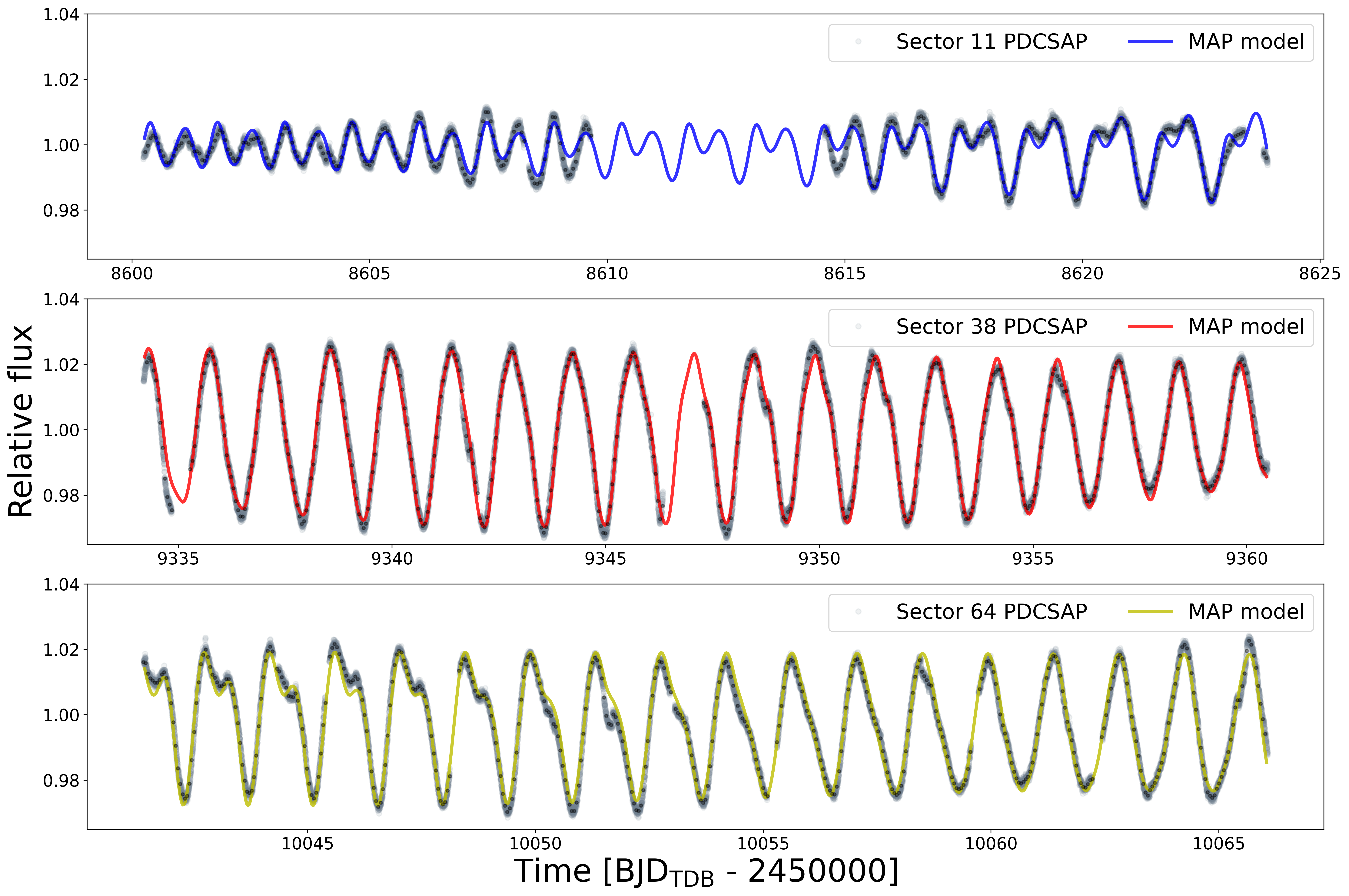}}
      \caption{Observed flux variability and \texttt{SAGE} model fits for HIP 67522. The morphology of the light curves changes strongly over time, which is best explained by the the evolution of active regions driven by the differential rotation of the star. The gray and black points represent individual and binned flux measurements from TESS sectors 11, 38 and 64, while the solid lines show the best-fit flux variability models that accounts for stellar differential rotation.}
         \label{fig: sage_model}
\end{figure*}

In addition to the occulted active regions, spots can also be present in regions outside the transit chord. Their effect is visible in the form of high-amplitude flux variability, which cannot be solely explained by the active regions we detected through spot occultations. To constrain the properties of active regions outside the transit chord, we modelled the stellar rotational variability using an Markov Chain Monte Carlo (MCMC) approach.

We obtained the stellar variability models using the \texttt{SAGE}\footnote{\url{https://github.com/chakrah/sage}} tool. As described in detail in \citet{sage}, \texttt{SAGE} creates a model light curve by rotating a stellar grid with active regions. The model light curve is parameterised by the position of active regions (their latitude and longitude), size, and contrast. We choose wide uniform priors on all parameters. The stellar inclination can be set as a free parameter, but we choose to fix it at 90$^{\circ}$ based on the constraints derived by \citet{heitzmann2021}, which find an inclination > 85$^{\circ}$ (3$\sigma$).

For HIP 67522, the morphology of both the TESS and CHEOPS light curves changes strongly over time, likely driven by the evolution of active regions on the stellar surface. In an associated paper by \citet{Loessnitz2025}, similar patterns was shown to arise from stable active regions combined with differential rotation of the star. Additionally, the same work warns against using the popular two-parameter equation for differential rotation, 

\begin{equation}
    \omega(\theta) = A + B \cdot \sin^2{\theta} 
\end{equation}

\noindent
where $\theta$ is the stellar latitude, $A$ is the equatorial rotation rate, and $B$ is the first-order shear factor. In solar physics, an additional corrective term $C. \sin^{4}{\theta}$ is often omitted because Sunspots predominantly form within $|\theta|<40^\circ$. For HIP 67522, however, we include the corrective term, to mitigate possible biases from higher latitude spots. To avoid introducing an additional parameter to fitting routines that are already difficult to constrain, we use the `frozen' two-parameter model proposed by \citet{Loessnitz2025}, where the parameters $B$ and $C$ are fixed to solar values and scaled as a whole by a new combined shear parameter $\beta$. The parameter $A$ remains the same, but is renamed $\alpha$ to avoid confusion. 

\begin{equation}
    \omega(\theta) = \alpha - \beta ( 2.504 \sin^2{\theta} + 2.23  \sin^4{\theta} ).\label{eq:2}
\end{equation}

\noindent
When $\beta$ equals unity, the star will have solar DR rates, a smaller value will result in a more rigid rotation, while a higher value indicates super-solar DR. 

We implemented Eq. \ref{eq:2} in \texttt{SAGE} to differentially rotate starspots, and assumed that the active regions are circular, uniform emitters, similar to $\alpha$ sunspots \citep{hale1919}. We modeled the TESS PDCSAP light curves with a 2-minute cadence. All transits of planet\,b and c were removed, along with any strong flux outliers and flares ($>$3$\times$median absolute deviation over a 15-point window). To reduce computational time, we binned the light curve into 500 uniformly spaced bins or a cadence of 1.13 hours. In addition, we chose to fix the spot temperatures to 5450 K to limit the degeneracies between spot size and temperature. This value corresponds to the average spot temperature derived from spot occultations observed in CHEOPS light curves (see Sect. \ref{subsec: spot_occultations}). We iteratively increased the number of spots from one to five and selected the model with the lowest BIC. We modelled all three sectors with three starspots. In Fig. \ref{fig: sage_model}, we present the TESS light curves, together with the best-fit models derived from the posteriors of our MCMC fit. The coefficients for differential rotation are reported in Table \ref{table: sage_differential_rotation}. For all three sectors, we constrained a super-solar shear of differential rotation.  

\begin{table}[h]
\centering
\setlength{\tabcolsep}{4pt}
      \caption[]{Fitted parameters for DR of HIP 67522}
         \label{table: sage_differential_rotation}
         \begin{tabular}{ccc}
            \hline
            \noalign{\smallskip}
            Sector & Equatorial rotation rate ($\alpha$) & shear of DR ($\beta$)  \\
             & [days] &  \\
            \noalign{\smallskip}
            \hline
            \noalign{\smallskip}
            S.11 (05-2019) & 1.413$\pm$0.002 & 1.4$\pm$0.2  \\
            \noalign{\smallskip}
            S.38 (05-2021) & 1.414$\pm$0.001 & 2.4$^{+4.6}_{-1.0}$ \\
            \noalign{\smallskip}
            S.64 (04-2023) & 1.413$^{+0.002}_{-0.003}$ & 2.2$^{+0.6}_{-0.3}$  \\
            \noalign{\smallskip}
            \hline
        \end{tabular}
\end{table}

\section{Modelling and Discussion}
\label{sec: discus}
\subsection{Dynamical modelling}\label{subsec: dynamical_modelling}

Measuring masses of young planetary systems is challenging due to the enhanced levels of stellar activity. \citet{dai2024} studied the occurrence rates of resonant planetary systems across different age groups and found a prevalence of resonant configurations in young planetary systems. HIP 67522\,b is near the first-order 2:1 resonance with the outer planet\,c. Thus, planet-planet gravitational interactions can result in deviations from the linear ephemeris of a strictly Keplerian orbit, leading to TTVs. 

The measured individual transit times do not show significant TTVs exceeding two minutes, nor do they show a clear periodicity. Nevertheless, we try to model them, as they can exclude some masses and orbital configurations that would result in a detectable signal. We modeled the transit timings using {\sc \tt TTVFast} \citep{Deck2014} and explored the parameter space with the nested‑sampling code {\sc \tt NAUTILUS} \citep{Lange2023}. We adopted a Gaussian prior for planet b's mass of $13.8\pm1.0$~M$_{\oplus}$ from \citet{Thao2024}, as well as Gaussian priors for the planet's orbital inclinations based on Table~\ref{table:orbital_solution}. Uniform priors were used for all remaining parameters. We obtained a 99\% upper limit for planet~c's mass of 22~M$_{\oplus}$ and an eccentricity of 0.181, assuming coplanarity (i.e., zero difference in longitude of the ascending node, $\Delta\Omega=0$) and 10~M$_{\oplus}$ and 0.221 eccentricity allowing for non-coplanarity (free $\Delta\Omega$ with a uniform prior between -90$^{\circ}$ and 90$^{\circ}$). The 99\% upper limits on the eccentricity of planet\,b are 0.142 under the coplanar assumption and 0.307 when non-coplanarity is allowed.

\subsection{Formation and evolution pathways}

   \begin{figure}
   \centering
\resizebox{\columnwidth}{!}{\includegraphics[trim=0.0cm 0.0cm -0.0cm 0.0cm]{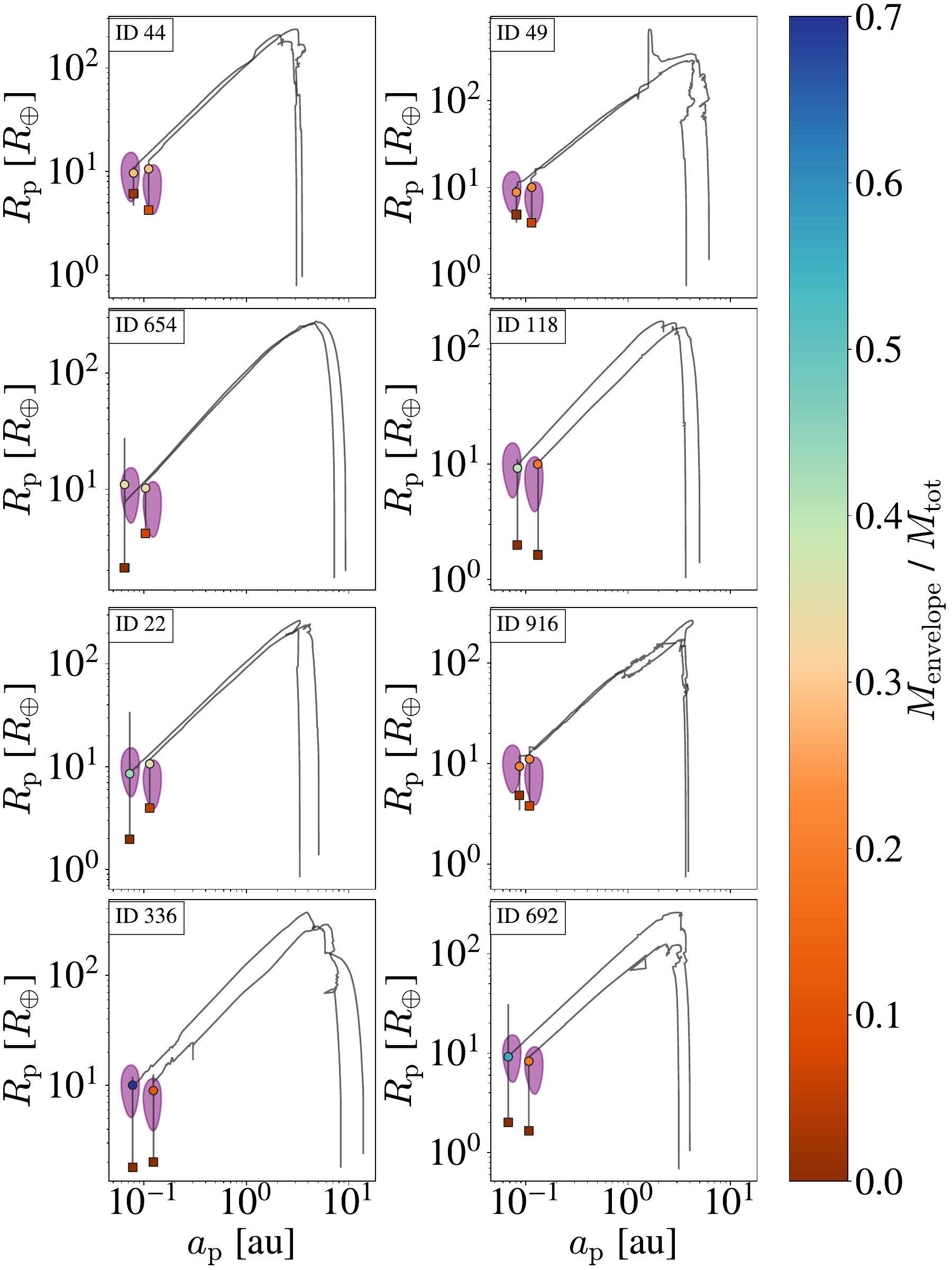}}
      \caption{Planetary radius versus semi-major axis for eight synthetic planetary systems analogues to HIP 67522 (one per panel, Bern model system ID shown in the top-left corner). Planets from the NGPPS population \citep{Emsenhuber2021a,Emsenhuber2021b} are shown at 20 Myr as circles and they are color-coded by envelope mass fraction. The purple ellipses indicate the selection constraints used to identify the analogues. Evolutionary tracks are shown as gray lines, with final planet locations and radii marked by squares, also color-coded by their final envelope fraction.}
         \label{fig: RvsSMA_colorcoded_envelope}
   \end{figure}

   \begin{figure}
   \centering
\resizebox{\columnwidth}{!}{\includegraphics[trim=0.0cm 0.0cm -0.0cm 0.0cm]{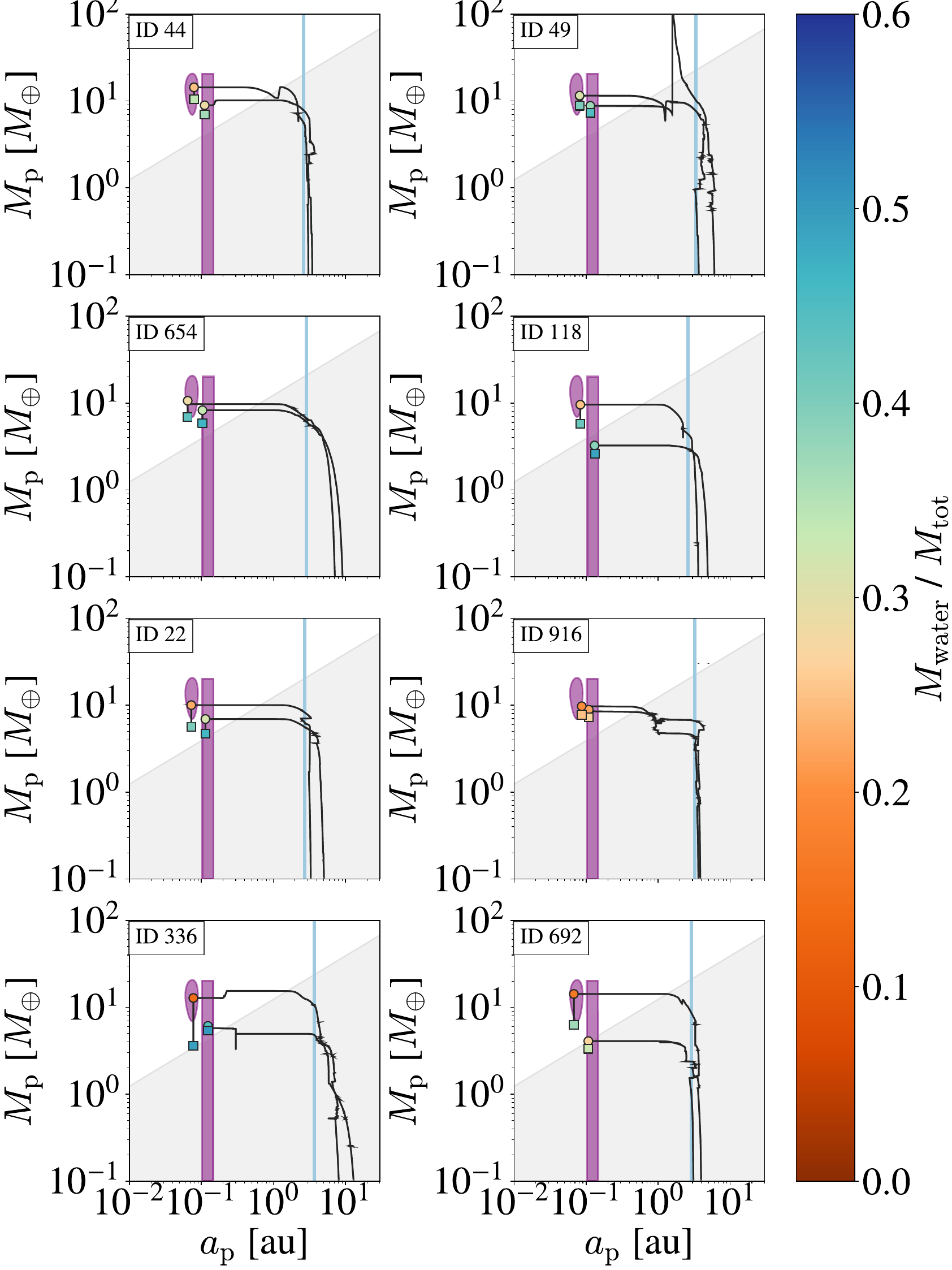}}
      \caption{Planetary mass versus semi-major axis for the same eight synthetic analogues to HIP 67522 shown in Fig. \ref{fig: RvsSMA_colorcoded_envelope}. Circles represent planets at 20 Myr, while squares indicate the final planet positions and masses. The color-bar shows the water mass fraction, while the purple rectangle and ellipse indicate the selection constraints use to identify the analogues. The shaded gray region marks the region below the RV detectability threshold of 1 m/s, while the formation tracks are shown as gray lines. The vertical blue line marks the initial location of the water ice line of each synthetic system.}
         \label{fig: MvsSMA_colorcoded_water}
   \end{figure}

To investigate possible formation and evolution pathways for the HIP 67522 system, we compare the observed two-planet architecture with synthetic planetary systems from the Bern model of planet formation and evolution \citep{Alibert2004,Alibert2005,Mordasini2012a,Mordasini2012b,Alibert2013,Fortier2013,Emsenhuber2021a,Emsenhuber2021b}. In particular, we use the NG76 population from the New Generation Planetary Population Synthesis (NGPPS; \citealt{Emsenhuber2021a,Emsenhuber2021b}), publicly available on \texttt{DACE}\footnote{\url{https://dace.unige.ch/populationAnalysis/?populationId=NG76}}. This population is tailored for solar-mass stars and simulates the growth of initially 100 randomly distributed lunar-mass embryos per disc.

During the formation phase, the model includes the accretion of planetesimals and gas onto the embryos, the evolution of the protoplanetary disc, gravitational interactions between forming planets, and disc-driven orbital migration. After disc dispersal, the long-term evolution of each planet is followed individually up to 10 Gyr, accounting for atmospheric photoevaporation of H/He envelopes, tidal migration, and the thermal evolution (cooling and contraction) of the planetary interior. We note that water is assumed to be accreted into the planetary core.

To identify viable formation and evolution pathways for HIP 67522\,b and c, we search for analogues of the observed system in the NG76 synthetic population at the age of 20 Myr, following a similar methodology to \citet{Ulmer-Moll2023} and \citet{Egger2024}. 
As a metric to identify analogues, we adopt the logarithmic distance in semi-major axis, as in \citet{KaufmannAlibert2023} and \citet{Egger2024}. Instead of comparing planet masses, we use planet radii, since the inner planet's mass is derived indirectly and the outer planet has only an upper limit on its mass. We define a synthetic planet as an analogue if it lies within an elliptical region in the semi-major axis–radius space, corresponding to a 20\% variation in the logarithm of the semi-major axis and a 50\% variation in the logarithm of the planet radius, relative to the observed values. These thresholds reflect both the limited size of the synthetic population (1000 systems) and the uncertainty in the observed planet radii due to stellar activity.

Furthermore, we require that both planets originate from the same synthetic system. After identifying initial analogues in the semi-major axis–radius space, we impose an additional constraint: the analogue to planet b must also lie within an elliptical region in the semi-major axis–mass space, allowing for a 20\% variation in log semi-major axis and a 50\% variation in log planet mass. For planet c, which has only an upper limit on the mass, we define a rectangular region that extends up to the observed mass limit and spans a 20\% variation in log semi-major axis.
Applying this multi-step selection procedure yields eight synthetic systems, illustrated in Figures \ref{fig: RvsSMA_colorcoded_envelope} and \ref{fig: MvsSMA_colorcoded_water}.

From the gray formation tracks in Fig.~\ref{fig: MvsSMA_colorcoded_water}, we find that all analogue planets form beyond the ice line. Through the accretion of planetesimals and collisions with other embryos in the system, the planets grow to several Earth masses and begin migrating inwards, often experiencing additional impacts along the way (e.g., systems ID 49, ID 336 and ID 916).

\begin{figure*}
    \centering
    \includegraphics[width=0.8\linewidth]{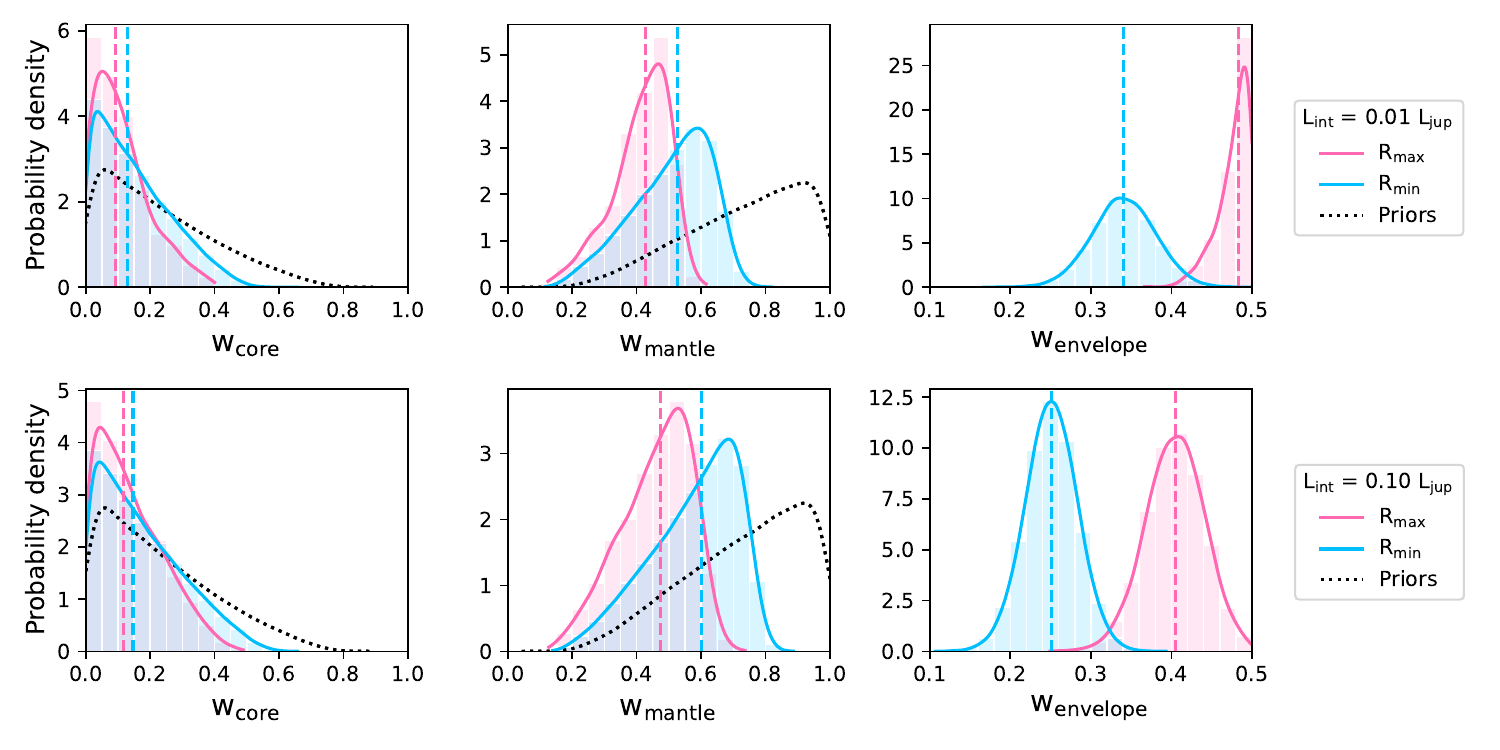}
    \caption{Inferred internal structure for HIP~67522\,b. Depicted are the mass fractions of the inner core, w$_\mathrm{core}$, the silicate mantle, w$_\mathrm{mantle}$, and the H/He dominated envelope, w$_\mathrm{envelope}$, both assuming the minimal (blue lines) and maximal planetary radius (pink lines) given the measured transit depths of the individual TESS and CHEOPS lightcurves. The top and bottom row show results for different intrinsic luminosity values, in line with the results of our formation and evolution analysis. The dotted black lines show the chosen priors, while the dashed vertical lines show median values of the posteriors.}
    \label{fig:int_struct_b}
\end{figure*}


By 20 Myr, the analogues have accreted H/He envelope—especially those corresponding to planet b, which have more than 20\% of their mass in H/He—and have incorporated water into their cores, consistent with their formation beyond the ice line. However, photo-evaporation eventually strips both planets of their envelopes during the subsequent evolution, leaving them with less than 15\% of H/He. Initially, planet b analogues have a higher envelope fraction than those of planet c, but both planets progressively lose their atmospheres, leading to a reduction in planetary radius over time. It is interesting to note that over 10 Gyr both planets could either evolve towards a Neptune-size planet, or towards an Earth-size planet with a completely stripped off atmosphere. The outcome primarily depends on the envelope fraction at 20 Myr, and in the case of planet c, also on the planetary mass. At 20 Myr, our analogues show that when planet c has a mass below 7 $M_{\oplus}$ and an envelope fraction $M_{\rm env}/M_{\rm tot} < 0.20$, it evolves towards $R_{\rm p} \lesssim 2\ R_{\oplus}$. Conversely, when planet b begins with $M_{\rm env}/M_{\rm tot} > 0.30$, it evolves to $R_{\rm p} \lesssim 2\ R_{\oplus}$ as well, largely independent of its total mass.

\subsection{Internal structure modeling}

As a next step, we used the \texttt{plaNETic} framework\footnote{\url{https://github.com/joannegger/plaNETic}} \citep{Egger2024} to infer the internal structure of HIP~67522\,b. \texttt{plaNETic} is based on the \texttt{BICEPS} planetary structure model \citep{Haldemann2024} and uses a neural network as a fast surrogate for this model in an accept-reject sampling scheme to infer posterior distributions of the planetary interior. The planet is modeled as an inner core made up of iron and up to 19\% of sulfur, a silicate mantle of oxidized silicon, magnesium and iron, and an envelope consisting of a mixture of hydrogen, helium, and water.

We sampled the planetary Si/Mg/Fe ratios uniformly from the simplex where their molar fractions add up to one, with an additional upper limit of 75\% on the iron fraction. We further assumed the envelope to consist of mostly H/He and only a very small water mass fraction. For further details on the model and the chosen priors, we refer to \citeauthor{Egger2024}~(\citeyear{Egger2024}, prior option B3).

While \texttt{plaNETic} usually uses the age-luminosity fit of \cite{Mordasini2020} to estimate the planet's intrinsic luminosity, HIP~67522 is younger than the planets this fit is applicable for. We therefore ran two separate models with intrinsic luminosities corresponding to the extreme values of the synthetic planetary analogues identified in the previous section, 1\% and 10\% of Jupiter's luminosity. To study the impact of the transit depth variations between the individual transit observations, we further ran two separate models representing the maximum and minimum radius value for HIP67522\,b.

The most important posterior distributions are presented in Figures~\ref{fig:int_struct_b}. For both intrinsic luminosity values, there is a clear difference between the inferred planetary envelope mass fractions for the minimal and maximal transit depth of planet~b, with envelope mass fractions of $34.1\pm3.9$\% (minimal radius) and $48.3^{+1.3}_{-2.5}$\% (maximal radius) for an assumed an intrinsic luminosity of 1\%~L$_\textrm{jup}$ and $25.1\pm3.2$\% (minimal radius) and $40.5^{+3.5}_{-3.6}$\% (maximal radius) for an intrinsic luminosity of 10\%~L$_\textrm{jup}$. To a lesser extent, also the corresponding posterior distributions of the mantle layer are distinguishable but still have significant overlap. 


\section{Conclusion}\label{sec:concl}

In this study, we analysed photometric transits of HIP 67522\,b and \,c using new data acquired from CHEOPS, TESS and multiple-ground-based observatories. Our key findings are summarised below:

\begin{enumerate}
    \item Transit timing variations and mass of planet\,c: We found no TTVs exceeding two minutes for the inner planet HIP 67522\,b. Using independent mass constraints from atmospheric retrievals of planet\,b and from photo-dynamical modelling, we derived an upper limit on the mass of planet\,c of 22 M$_\oplus$ when assuming co-planarity with the inner planet and 10 M$_\oplus$ when allowing for different inclinations. This low mass estimate makes the system a compelling target for further atmospheric characterisation with transmission spectroscopy. 

    \item Transit depth variations: Using TESS and CHEOPS transits of HIP 67522\,b, we detected transit depth variations with amplitudes exceeding 30\%. The inferred planetary radius varies from roughly 8 to 12 R$_{\oplus}$, highlighting the significant biases introduced by stellar activity when measuring the radii of transiting planets around young and active stars. 

    \item Stellar activity and rotation: By combining observations acquired over multiple years, we measured a stellar rotational period of 1.4262$\pm$0.0073 days, derived from rotation-induced flux variability. This period is nearly commensurate, 5:1 and 10:1, with the orbital periods of HIP 67522\,b and c. Furthermore, the amplitude and overall shape of the out-of-transit flux modulation vary significantly over time. We show that this variability arises from the evolution of active regions on the stellar surface, driven by differential rotation. By modelling the flux variability, we constrained the differential rotation rate of HIP 67522 to be supersolar.

    \item Formation and evolution pathways: We find that both planets\,b and c likely formed beyond the water ice-line and migrated inward, accreting substantial H/He envelopes during the process. For planet\,b, at an age of 20 Myr, more than 20\% of its total mass is incorporated in the atmospheric envelope. Due to photoevaporation, the planets will gradually lose their atmospheres, eventually evolving into either Earth size planets with completely stripped atmospheres or sub-Neptune-sized planets, largely depending on the envelope mass fraction at 20 Myr. 

    \item Internal structure: For HIP 67522 b, we inferred an interior composition dominated by a substantial gaseous envelope, considering both a high and a low intrinsic luminosity case. We also find that the detected large transit depth variations translate to similarly large variations in the inferred envelope mass fractions between 30 and 40\%, depending on the assumed intrinsic luminosity of the planet. A smaller variation in the inferred mantle mass fractions is also noticeable, but the corresponding posteriors still mostly overlap.    
\end{enumerate}


\begin{acknowledgements}
HC, ML, MPB, BA and DJMR acknowledge support of the Swiss National Science Foundation under grant number PCEFP2\_194576. This work has been carried out within the framework of the NCCR PlanetS supported by the Swiss National Science Foundation under grants 51NF40\_182901 and 51NF40\_205606. AN and JV acknowledges support from the Swiss National Science Foundation (SNSF) under grant PZ00P2\_208945. JAE acknowledges support through the European Space Agency (ESA) Research Fellowship Programme in Space Science. MPB and EG acknowedlges support from UK Research and Innovation (UKRI) under the UK government's Horizon Europe funding gurantee [grant number EP/Z000890/1]. AP is supported by the \emph{Deut\-sche For\-schungs\-ge\-mein\-schaft, DFG\/} project number PI 2102/1-1. ADF acknowledges funding from NASA through the NASA Hubble Fellowship grant HST-HF2-51530.001-A awarded by STScI. EI acknowledges funding from the European Research Council under the European Union's Horizon Europe program (grant number 101042416 STORMCHASER). SS acknowledges Fondo Comit´e Mixto-ESO Chile ORP 025/2022. JSJ greatfully acknowledges support by FONDECYT grant 1240738 and from the ANID BASAL project FB210003. The work of HPO has been carried out within the framework of the NCCR PlanetS supported by the Swiss National Science Foundation under grants 51NF40\_182901 and 51NF40\_205606.
CHEOPS is an ESA mission in partnership with Switzerland with important contributions to the payload and the ground segment from Austria, Belgium, France, Germany, Hungary, Italy, Portugal, Spain, Sweden, and the United Kingdom. The CHEOPS Consortium would like to gratefully acknowledge the support received by all the agencies, offices, universities, and industries involved. Their flexibility and willingness to explore new approaches were essential to the success of this mission. CHEOPS data analysed in this article will be made available in the CHEOPS mission archive (\url{https://cheops.unige.ch/archive_browser/}). This paper includes data collected by the TESS mission. Funding for the TESS mission is provided by the NASA’s Science Mission Directorate. The computations in this paper were performed at the University of Geneva on the ``Baobab” HPC cluster. 
\end{acknowledgements}

%
\bibliographystyle{aa} 
\bibliography{hip67522}

\begin{appendix}
\onecolumn
\section{Additional Tables}
\begin{table}[htb]
    \caption{CHEOPS observations log}
    \centering
    \begin{tabular}{@{}ccccl@{}}
        \toprule
         & File Key & Obs Start & Obs Stop & Remarks\\
        \midrule
        0 & CH\_PR240004\_TG000101\_V0300 & 2024-03-09T05:09:00 & 2024-03-09T18:20:00 & Partial transit\\
        1 & CH\_PR240017\_TG000801\_V0300 & 2024-03-11T08:35:54 & 2024-03-12T02:25:00 & Stellar rotation\\
        2 & CH\_PR240004\_TG000102\_V0300 & 2024-03-16T05:56:00 & 2024-03-16T19:07:00 & Stellar rotation\\
        3 & CH\_PR240017\_TG000901\_V0300 & 2024-03-21T03:03:00 & 2024-03-21T19:45:00 & Stellar rotation\\
        4 & CH\_PR240004\_TG000103\_V0300 & 2024-03-23T03:34:00 & 2024-03-23T16:45:00 & Partial transit\\
        5 & CH\_PR240004\_TG000104\_V0300 & 2024-03-30T02:46:59 & 2024-03-30T15:58:00 & Partial transit*\\
        6 & CH\_PR240017\_TG000101\_V0300 & 2024-04-05T15:37:56 & 2024-04-06T11:37:00 & Full transit \\
        7 & CH\_PR240017\_TG000601\_V0300 & 2024-04-11T10:48:00 & 2024-04-12T02:20:00 & Stellar rotation\\
        8 & CH\_PR240017\_TG000102\_V0300 & 2024-04-12T14:38:55 & 2024-04-13T09:45:00 & Full transit\\
        9 & CH\_PR240017\_TG000103\_V0300 & 2024-04-19T13:41:55 & 2024-04-20T12:28:00 & Full transit\\
        10 & CH\_PR240017\_TG000501\_V0300 & 2024-04-28T17:25:00 & 2024-04-29T08:57:00 & Stellar rotation\\
        11 & CH\_PR240017\_TG000701\_V0300 & 2024-04-30T12:18:55 & 2024-05-01T03:51:00 & Stellar rotation\\
        12 & CH\_PR240017\_TG000104\_V0300 & 2024-05-03T12:11:52 & 2024-05-04T07:29:00 & Full transit \\
        13 & CH\_PR240004\_TG000105\_V0300 & 2024-05-10T20:52:53 & 2024-05-11T10:41:00 & Partial transit\\
        14 & CH\_PR240004\_TG000106\_V0300 & 2024-05-17T19:31:53 & 2024-05-18T08:43:00 & Partial transit\\
        15 & CH\_PR240004\_TG000107\_V0300 & 2024-05-24T17:53:52 & 2024-05-25T08:56:00 & Partial transit*\\
        16 & CH\_PR240017\_TG001001\_V0300 & 2024-05-26T15:30:00 & 2024-05-27T08:09:00 & Stellar rotation\\
        17 & CH\_PR240004\_TG000108\_V0300 & 2024-05-31T18:07:00 & 2024-06-01T07:18:00 & Partial transit\\
        18 & CH\_PR240004\_TG000109\_V0300 & 2024-06-07T16:58:00 & 2024-06-08T06:09:00 & Partial transit*\\
        19 & CH\_PR240004\_TG000110\_V0300 & 2024-06-14T17:24:00 & 2024-06-15T07:34:00 & Stellar rotation\\
        20 & CH\_PR240004\_TG000111\_V0300 & 2024-06-21T14:03:00 & 2024-06-22T03:14:00 & Partial transit\\
        \bottomrule
    \noalign{\smallskip}
    \multicolumn{5}{l}{
    \begin{minipage}{0.8\linewidth}
    \small
    \textit{Notes:} * marks partial transits observations with no ingress or egrees visibility to reliability constrain transit times. 
    \end{minipage}
    }
    \end{tabular}

    \label{tab: CHEOPS_logs}
\end{table}

\begin{table}[htb]
    \caption{Transit times and planet-to-star radius ratio for CHEOPS, TESS and NGTS observations of planet\,b}
    \centering
    \begin{tabular}{lll}
        \toprule
          Transit & $T_{0}$ [BJD$_{\rm{TDB}}$] & $R_{\rm{pl}}/R_{\star}$ \\
        \midrule
         S11.1 & 2458597.0652$\pm$0.0007 & 0.070$\pm$0.001\\
         S11.2 & 2458604.0240$\pm$0.0006 & 0.067$\pm$0.001\\
         S11.3 & 2458610.982$\pm$0.001 & 0.069$\pm$0.001\\
         S11.4 & 2458617.9435$\pm$0.0008 & 0.061$\pm$0.002\\
         \noalign{\smallskip}
         S38.2 & 2459341.7284$\pm$0.0006 & 0.078$\pm$0.001\\
         S38.3 & 2459348.6878$\pm$0.0005 & 0.070$\pm$0.001\\
         S38.4 & 2459355.6456$\pm$0.0009 & 0.060$\pm$0.003\\
         \noalign{\smallskip}
         NGTS 1 & 2459703.617$\pm$0.001 & N/A \\
         NGTS 2 & 2459710.577$\pm$0.002 & N/A \\
         NGTS 3 & 2459724.450$\pm$0.002 & N/A \\
         \noalign{\smallskip}
         S64.1 & 2460044.635$\pm$0.001 & 0.056$\pm$0.003\\
         S64.2 &2460051.5931$\pm$0.0007 & 0.068$\pm$0.001\\
         S64.3 & 2460058.5541$\pm$0.0006 & 0.066$\pm$0.001\\
         S64.4 & 2460065.5132$\pm$0.0007 & 0.074$\pm$0.002\\
         \noalign{\smallskip}
         CH\_PR240004\_TG000101\_V0300 & 2460378.688$\pm$0.001 & N/A \\
         CH\_PR240004\_TG000103\_V0300 & 2460392.608$\pm$0.001 & N/A \\
         CH\_PR240017\_TG000101\_V0300 & 2460406.5271$\pm$0.0006 & 0.0799$\pm$0.0006 \\
         CH\_PR240017\_TG000102\_V0300 & 2460413.487$\pm$0.001 & 0.060$\pm$0.001 \\
         CH\_PR240017\_TG000103\_V0300 & 2460420.448$\pm$0.001 & 0.0639$\pm$0.0006 \\
         CH\_PR240017\_TG000104\_V0300 & 2460434.3659$\pm$0.0007 & 0.068$\pm$0.001 \\
         CH\_PR240004\_TG000105\_V0300 & 2460441.324$\pm$0.001 & N/A \\
         CH\_PR240004\_TG000106\_V0300 & 2460448.282$\pm$0.001 & N/A \\
         CH\_PR240004\_TG000108\_V0300 & 2460462.2049$\pm$0.0009 & N/A \\
         CH\_PR240004\_TG000111\_V0300 & 2460483.081$\pm$0.001 & N/A \\
        \bottomrule
    \end{tabular}
    \label{tab: transit_times_radius_ratio}
\end{table}

\begin{table}[!htb]
    \caption{Transit times for TESS observations of planet\,c}
    \centering
    \begin{tabular}{ll}
        \toprule
          Transit & $T_{0}$ [BJD$_{\rm{TDB}}$] \\
        \midrule
         S11.1 & 2458602.502$\pm$0.001 \\
         S11.2 & 2458616.839$\pm$0.002 \\
         \noalign{\smallskip}
         S38.1 & 2459347.915$\pm$0.001 \\
         \noalign{\smallskip}
         S64.1 & 2460050.326$\pm$0.001 \\
         S64.2 & 2460064.663$\pm$0.002 \\
        \bottomrule
    \end{tabular}
\end{table}
\newpage
\section{Additional Figures}

\begin{figure*}
   \centering
\resizebox{\textwidth}{!}{\includegraphics[trim=0.0cm 0.0cm 0.0cm 0.0cm]{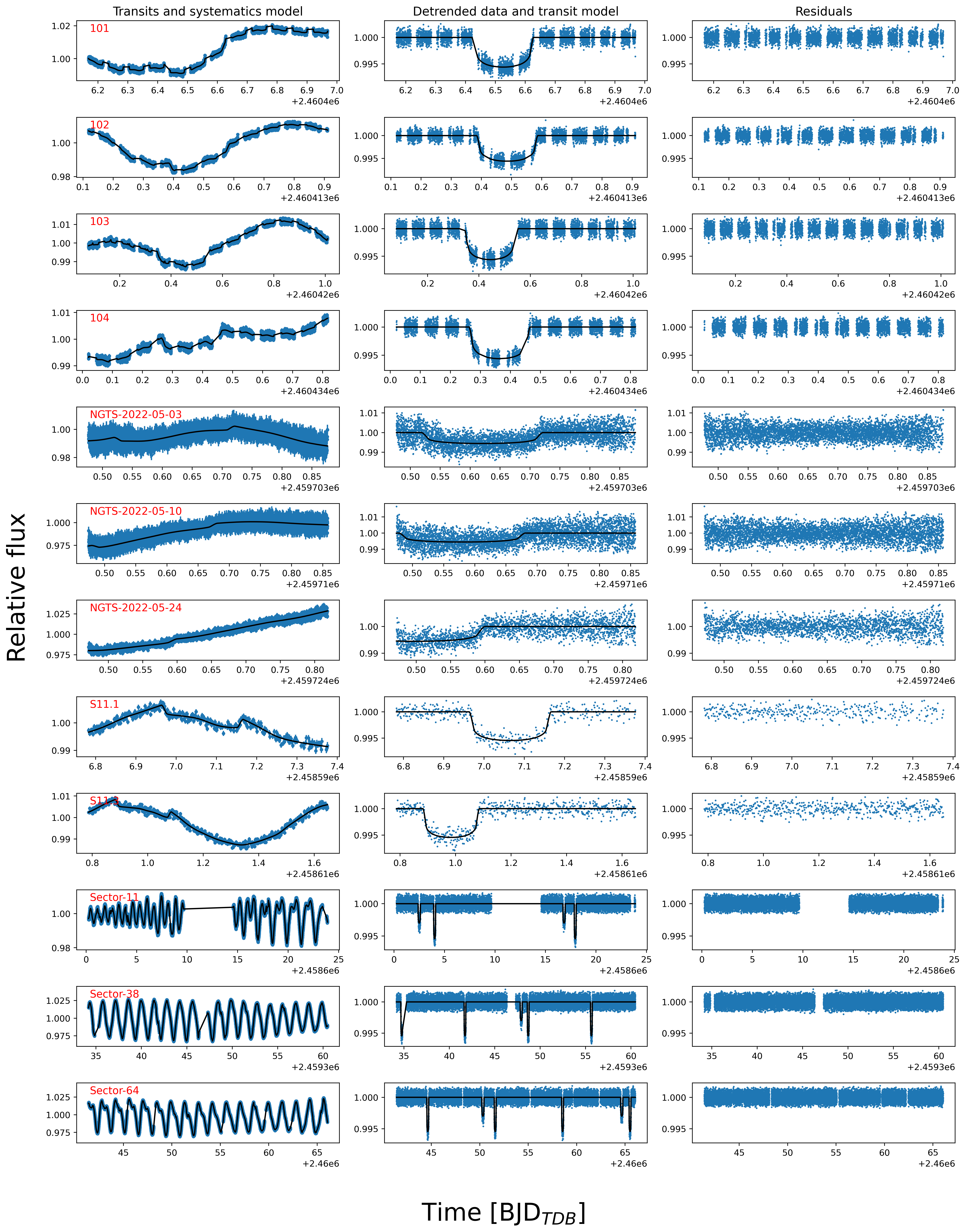}}
      \caption{CHEOPS, NGTS and TESS light curves of HIP 67522\,b and c. \emph{Left:} The best-fit transit and systematics model over-plotted over raw data. \emph{Middle:} Pure transit model overplotted over detrended data. \emph{Right:} Residuals for each observation. }
         \label{fig:cheops_ngts_tess_lightcurves}
\end{figure*}

\begin{figure*}
   \centering
\resizebox{\textwidth}{!}{\includegraphics[trim=0.0cm 0.0cm 0.0cm 0.0cm]{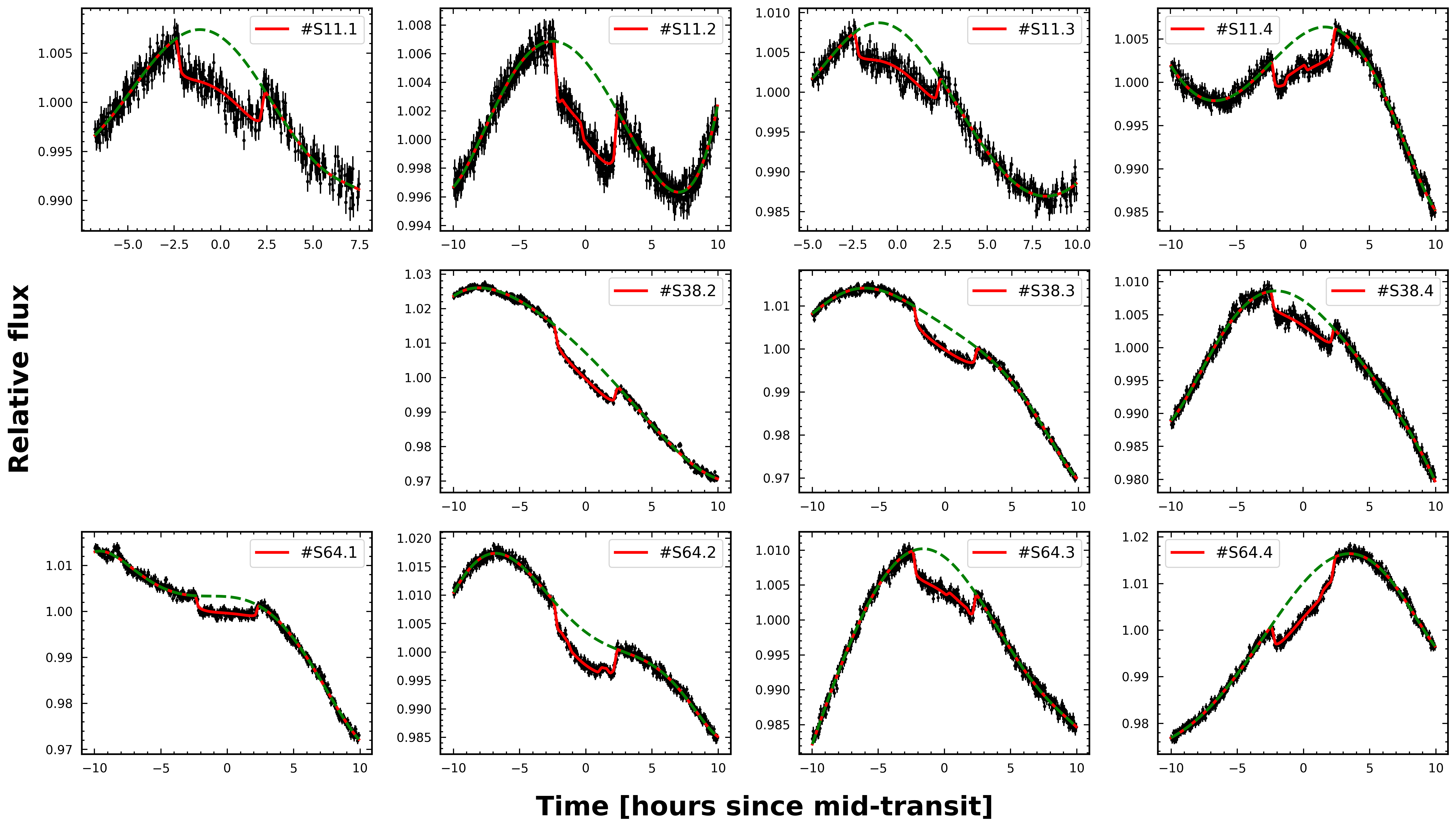}}
      \caption{TESS light curves of HIP 67522\,b. The best-fit transit model with spot occultations are overplotted on top. The transit 38.1 is removed due to strong flare contamination.}
         \label{fig:photLightcurve}
\end{figure*}

\begin{figure}
   \centering
\resizebox{0.5\textwidth}{!}{\includegraphics[trim=0.0cm 0.0cm -0.0cm 0.0cm]{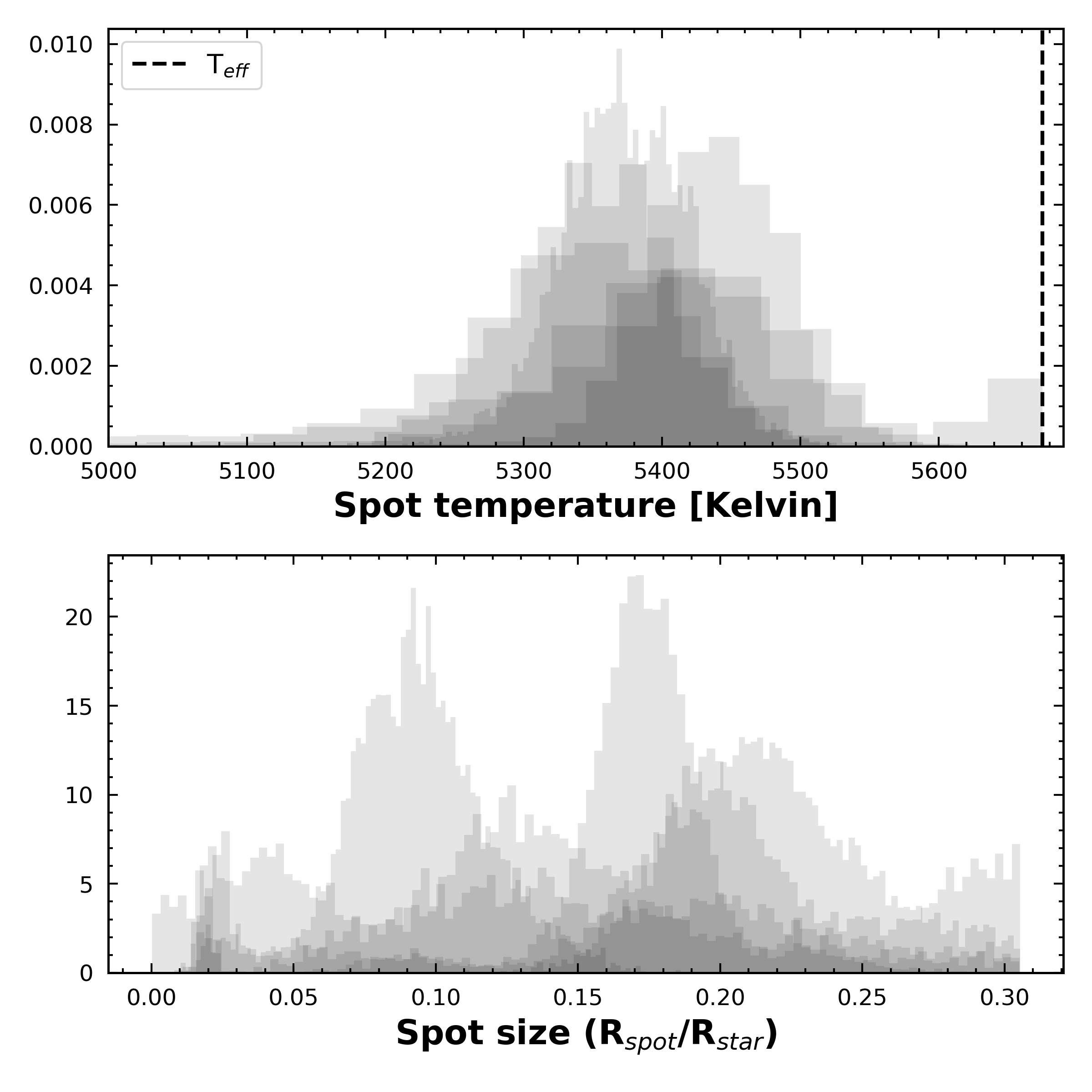}}
      \caption{Posteriors of spot temperature (\emph{top}) and size (\emph{bottom}) for the TESS-detected spot occultations of HIP 67522 shown in Fig.~\ref{fig:photLightcurve}.  The dashed vertical line on the top panel is the effective temperature of the star.}
         \label{fig: tess_spot_temp}
\end{figure}



\end{appendix}

\end{document}